\documentclass{aa}

\usepackage{graphicx}
\usepackage{txfonts}
\usepackage{amsmath}
\usepackage{hyperref}
\usepackage{natbib}
\usepackage{placeins}

\begin{document}

\makeatletter
\let\linenumbers\relax
\let\linenumbersep\relax
\let\resetlinenumber\relax
\makeatother

\title{Characterizing Low-Latency Sky Localization in Multi-Detector Gravitational-Wave Networks}

\author{Amazigh Ouzriat\inst{1}\fnmsep\thanks{Corresponding author.}
        \and Viola Sordini\inst{1}
        \and Francesco Di Renzo\inst{2,3}
}

\institute{Institut de Physique des 2 Infinis de Lyon (IP2I), CNRS/IN2P3, Université Claude Bernard Lyon 1, Villeurbanne, France\\
          \email{a.ouzriat@ip2i.in2p3.fr}
      \and Università di Firenze, Sesto Fiorentino I-50019, Firenze, Italy
      \and INFN, Sezione di Firenze, Sesto Fiorentino I-50019, Firenze, Italy\\
}


\abstract
{Low-latency analyses of gravitational-wave (GW) data from LIGO, Virgo, and KAGRA enable rapid detection of compact binary coalescences (CBC) and prompt sky localization, essential for electromagnetic follow-up in multi-messenger astronomy.}
{We evaluate the performance and limitations of low-latency sky localization using BAYESTAR algorithm, and investigate the impact of low-significance Virgo triggers.}
{We inject simulated CBC signals into Gaussian-stationary noise and into Virgo data from the second part of the third LIGO-Virgo observing run (O3b), then reconstruct skymaps across multiple detector network configurations. Localization accuracy is assessed using Percentile-Percentile plots, the Jaccard index, and the Kullback-Leibler divergence.}
{Binary neutron star mergers are statistically consistent with ideal calibration, showing deviations below $3\sigma$, particularly when Virgo is included in the network, whereas skymaps for neutron star–black hole and binary black hole mergers tend to be overconfident. Adding a third detector generally improves accuracy, but the searched area can degrade when Virgo's signal-to-noise ratio is low (SNR $\leq$ 5). For high-SNR events, relying on two detectors can mislocalize the source. Excluding Virgo can therefore cause the HL skymap to miss the true location when Virgo has strong antenna response; in such cases a three-detector configuration is required to recover the correct position and avoid misleading multi-messenger follow-up.}
{We introduce diagnostics to flag problematic skymaps and apply them to O3 public alerts, recovering simulation-predicted trends and flagging a few anomalous morphologies. The results are relevant for improving rapid vetting of GW alerts and guiding observational strategies in multi-messenger astronomy.}

\keywords{gravitational waves -- methods: data analysis -- methods: statistical -- techniques: interferometric}

\maketitle

\section{Introduction}

Compact binary coalescences (CBCs) are the primary sources observed by the current network of ground-based gravitational-wave (GW) detectors: LIGO Hanford and LIGO Livingston \citep{LIGOScientific:2014pky}, Virgo \citep{VIRGO:2014yos}, and KAGRA \citep{KAGRA:2020tym}. Low-latency pipelines such as \texttt{Aframe} \citep{Marx:2024wjt}, \texttt{cWB} \citep{Drago:2020kic}, \texttt{GstLAL} \citep{Cannon:2020qnf}, \texttt{MBTA} \citep{Aubin:2020goo, Allene:2025saz}, \texttt{PyCBC Live} \citep{Nitz:2018rgo, DalCanton:2020vpm}, and \texttt{SPIIR} \citep{Chu:2020pjv} continuously analyze the data in real time to identify GW candidates. Among these, \texttt{Aframe} is a machine-learning–based pipeline, while \texttt{cWB} performs a \emph{weakly modelled} coherent burst search, identifying excess coherent power with minimal assumptions on the GW waveform morphology. \texttt{GstLAL}, \texttt{MBTA}, \texttt{PyCBC Live}, and \texttt{SPIIR} are \emph{modelled} searches targeting CBCs. Once a significant candidate is detected, a public alert is issued with notices and circulars via the General Coordinates Network (GCN) \citep{GCN} and SCIMMA \citep{Brazier2021SCIMMA}. These alerts enable timely electromagnetic (EM) follow-up by the broader multi-messenger astronomy community, whereby telescopes and observatories across the EM spectrum (radio, optical, X-ray, gamma-ray) are directed toward the most probable sky region to search for transient counterparts to the GW event.

Following detection, the sky position of the source is estimated using the \texttt{BAYESTAR} algorithm \citep{Singer2016_BAYESTAR}, which performs rapid Bayesian inference based on triangulation from signal arrival times, further constrained by amplitude, phase, and signal-to-noise ratio (SNR) consistency across the detector network \citep{Fairhurst:2009tc}. The algorithm infers the joint posterior distribution for the source’s sky location -- right ascension ($\alpha$) and declination ($\delta$) -- and luminosity distance ($d_L$). The resulting skymap, typically available within tens of seconds, is critical for enabling prompt EM follow-up. While a two-detector network can constrain the sky position, the inclusion of a third detector substantially improves accuracy, particularly when it provides a long baseline and complementary orientation. A globally distributed network enhances angular resolution and reduces degeneracies in the inferred sky location.

Since the first CBC detection in 2015 \citep{LIGOScientific:2016aoc} during the initial LIGO observing run (O1), the global GW detector network has expanded. The Virgo interferometer joined the network in the second observing run (O2), providing additional triangulation capabilities. A landmark demonstration of this three-detector configuration was \texttt{GW170817} \citep{LIGOScientific:2017vwq}, a binary neutron star merger localized to $\sim$28 square degrees, which led to the first detection of an EM counterpart to a GW event. The Japanese detector KAGRA is actively participating (although with low sensitivity) to the third part of the fourth observing run (O4c), and the future addition of LIGO-India will further improve global baseline coverage. GW detections accumulated from O1 through the first part of the fourth observing run (O4a) culminate in a total of 218 confident CBC events \citep{LIGOScientific:2025slb}.

Modelled low-latency pipelines are based on a technique called \textit{matched filtering} that involves correlating the detector data with a bank of precomputed waveform templates representing expected GW signals from CBCs. These templates span a discrete set of parameters (such as masses and spins), meaning that detections are often triggered by the template that best matches the data, even if it does not exactly correspond to the true parameters of the source. \citep{Duverne:2023joq} show that even a minor mismatch between the true intrinsic parameters of a signal and those reported by the search pipeline, could induce an overestimation of the sky localization uncertainties. To account for this effect, a multiplicative correction factor, denoted $\xi$, is introduced to rescale the likelihood used in rapid Bayesian inference. The value of $\xi$ is pipeline-dependent because of differences in template banks. During the third observing run, a common value of $\xi=0.83$ -- originally optimized for \texttt{GstLAL} \citep{Singer2016_BAYESTAR} -- was applied across all pipelines to account for localization uncertainties. \citet{Duverne:2023joq} showed that $\xi=0.87$ is optimal for \texttt{PyCBC Live}. The same study also demonstrated that such correction is unnecessary (i.e. setting $\xi = 1$) and yields properly normalized posteriors when the waveform used in the analysis exactly matches the injected signal. While the skymap posteriors were generally consistent with expectations, residual deviations from perfect calibration remained, particularly in the BBH population.

In low-latency searches, when two or more interferometers are observing, GW event candidates are constructed from coincident triggers of the same waveform template in each detector. During O3, Virgo’s sensitivity was typically lower than that of the LIGO detectors, often yielding SNRs around $\sim$ 2--3 for CBC signals \citep{KAGRA:2021vkt}. At such low SNR, there is a risk of spurious associations between true LIGO signals and noise fluctuations in Virgo. Understanding the impact of low-SNR data from Virgo on sky localization is essential for assessing the reliability of low-latency alerts and optimizing electromagnetic follow-up.

Moreover, contamination of data from instrumental or environmental noise transients (\textit{glitches}) can affect the inferred sky localization. Prior to issuing a circular for the GW alert, the LIGO-Virgo-KAGRA (LVK) collaboration carries out an event vetting procedure to assess the astrophysical origin of the candidate and to identify any potential transient noise contamination. If a data quality issue is identified near the time of merger, noise subtraction is applied to mitigate the glitch, and updated skymaps are subsequently distributed on longer timescales.

These considerations motivate the present study. We aim to investigate the performance of low-latency GW sky localization. Particular attention is given to the role of the Virgo detector, whose contribution depends critically on its sensitivity and orientation relative to the source location. The goal is to develop robust statistical criteria that can be used to flag potential issues in sky localization and improve the interpretation of low-latency GW skymaps. This methodology will then be applied to assess the sky localization performance of alerts issued during the third LIGO-Virgo observing run (O3).
\section{Methodology}
We conduct a simulation campaign of GW signals (\textit{injections}) to assess the performance and reliability of low-latency GW sky localization using \texttt{BAYESTAR} across varying detector network configurations and CBC source parameters. We specifically evaluate the impact of including Virgo under both stationary-Gaussian and realistic noise conditions. To this end, we make use of statistical tests to quantify skymap consistency and evaluate localization performance. We further measure the improvement in credible area resulting from the addition of a third detector to a baseline two-detector network.

\subsection{Simulation Framework}
We simulate astrophysical populations for three CBC categories: binary black hole (BBH), neutron star–black hole (NSBH), and binary neutron star (BNS) mergers. Source-frame parameters are drawn from models informed by the population analysis of the third release of the Gravitational-Wave Transient Catalog (GWTC--3) \citep{LIGOScientific:2021psn}. We use the \texttt{IMRPhenomXAS} waveform model \citep{Pratten:2020fqn}, a non-precessing phenomenological approximant calibrated to numerical relativity, chosen for its accuracy and computational efficiency in large-scale injection studies.

Injections are then processed by a tool available in \texttt{PyCBC}, called \texttt{pycbc\_make\_skymap}\footnote{\url{https://github.com/gwastro/pycbc/blob/master/bin/pycbc_make_skymap}}. For each injection, it generates fake strain data by adding the GW signal to simulated noise in each detector, builds a template using the injection parameters, performs matched filtering, and computes the SNR timeseries. The results are then passed to \texttt{BAYESTAR} to produce a posterior distribution over sky position and luminosity distance. We set the likelihood scaling factor $\xi = 1$ throughout this study and skymaps are produced for each possible combination of instruments: HL, LV, LV, HLV.

\begin{figure}[htbp]
\centering
\includegraphics[width=0.48\textwidth]{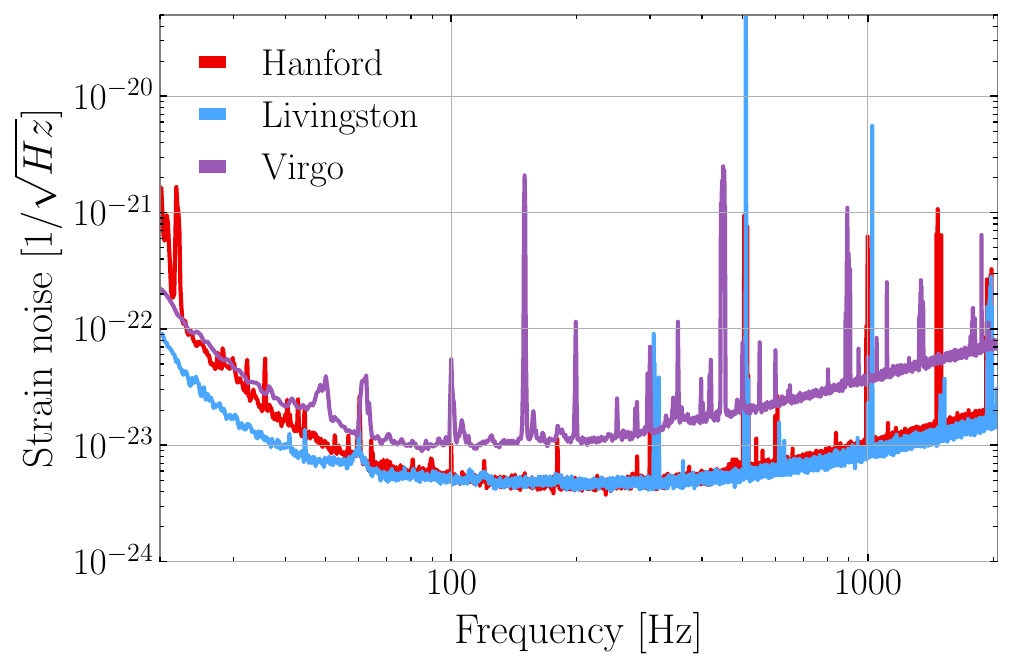}
\caption{Median amplitude spectral densities for LIGO Hanford, LIGO Livingston, and Virgo during the second part of the third observing run (O3b).}
\label{fig:o3_asds}
\end{figure}

\paragraph{Injection Selection.}  
We first implement a custom Python framework built on \texttt{LALSuite} \citep{lalsuite} and \texttt{PyCBC} to select injections likely to be detectable by the LIGO-Virgo network. For each source, the waveform is projected into the detector frame and matched-filtered against a single template with the exact same parameters as the injection, over the frequency band 20–1024 Hz. The maximum SNR is computed within a $\pm 50$ ms window around the injected coalescence time. All detectors use stationary Gaussian noise generated from representative amplitude spectral densities\footnote{\url{https://www.gw-openscience.org/GWTC-3/}} (ASDs) of the second part of the third LIGO-Virgo observing run (O3b), as shown in Fig.~\ref{fig:o3_asds}. 

We retain only injections satisfying an injected network SNR $\rho_{\mathrm{net}} > 8$ and at least one individual detector SNR $\rho_i > 5$, resulting in a set of $\sim$30,000 injections (10,000 per CBC category). 

\paragraph{Simulation Strategy.}  
We want to assess the impact of Virgo under both ideal stationary Gaussian noise and realistic noise conditions. We rerun the same set of 30,000 injections under each scenario, enabling a direct comparison of their respective effects. To isolate more subtle biases unrelated to major data quality issues, we restrict our analysis to Virgo data segments free from significant glitches.

The injection campaign proceeds in two steps:
\begin{enumerate}
    \item \textbf{Real-noise run:} LIGO Hanford and Livingston are simulated with Gaussian noise using representative O3b ASDs, while Virgo uses real O3b strain data from periods identified as clean of significant transient noise. To ensure glitch-free conditions, we use the Omicron pipeline -- a wavelet-based tool that detects excess power in the time-frequency domain, used for glitch identification -- to exclude any injection for which an Omicron glitch with SNR $> 6.5$ is present within $\pm1$s of the coalescence time.
    
    \item \textbf{Gaussian-noise run:} A second analysis is performed using stationary Gaussian noise in all detectors. The Virgo ASD employed in this run is not the O3b median ASD but the empirical amplitude spectral density locally estimated from the Virgo data segments used in the first run for the considered injection. This ensures that both runs assume identical Virgo sensitivity for each event.
\end{enumerate}

\subsection{Injection Distributions}

The injection parameters are drawn from astrophysically motivated distributions consistent with population models inferred from GWTC--3 \citep{LIGOScientific:2021psn}. Table~\ref{tab:injection_settings} summarizes the mass and spin distributions used for each CBC category.

For BBHs and the black hole component of NSBH systems, the primary mass $m_1$ is drawn from a power-law with a Gaussian peak:
\begin{equation}
    p(m_1) \propto m_1^{-\alpha} + \lambda \exp\left(-\frac{(m_1 - \mu)^2}{2\sigma^2}\right),
\end{equation}
where we use the GWTC--3 inferred values: $\alpha = 2.63$, $\lambda = 0.03$, $\mu = 33.1\,M_\odot$, and $\sigma = 5.69\,M_\odot$. The secondary mass is drawn using a mass ratio distribution $p(q) \propto q$, with $q = m_2/m_1 \leq 1$.

Neutron star masses for NSBH and BNS systems follow a truncated normal distribution centered at $1.33\,M_\odot$ with a standard deviation of $0.09\,M_\odot$, truncated to the interval $[1.0, 2.3]\,M_\odot$. Spins are aligned or anti-aligned with the orbital angular momentum and uniformly distributed in magnitude within the specified limits; spin precession is neglected in this study. Injections are distributed in redshift up to $z = 1.0$ ($D_L \approx 7\,\mathrm{Gpc}$) for BBH sources, and up to $z = 0.1$ ($D_L \approx 500\,\mathrm{Mpc}$) for NSBH and BNS systems. 

\begin{table*}[htbp]
\caption{Astrophysical distributions used to generate the injection populations for each CBC category.}
\label{tab:injection_settings}
\centering
\begin{tabular}{c c c c}
\hline
Category & Distribution & Mass range ($M_\odot$) & Spin range \\
\hline\hline
BBH & $p(m_1)$: PL+peak,\ $p(q) \propto q$ & $m_1 \in [5, 100],\ m_2 \leq m_1$ & $|\chi_{1,2}| \leq 1.0$ (uniform) \\
NSBH & $p(m_\mathrm{BH})$: PL+peak,\ $p(m_\mathrm{NS})$: truncated normal & $m_\mathrm{BH} \in [5, 100],\ m_\mathrm{NS} \in [1.0, 2.3]$ & BH: $|\chi| \leq 1.0$, NS: $|\chi| \leq 0.05$ \\
BNS & $p(m_{1,2})$: truncated normal & $m_{1,2} \in [1.0, 2.3]$ & $|\chi_{1,2}| \leq 0.05$ (uniform) \\
\hline
\end{tabular}
\end{table*}

\subsection{Statistical Tests}

To assess differences between sky localization posteriors reconstructed under varying detector networks or noise conditions, we employ three complementary diagnostics: PP plots \citep{Singer2016_BAYESTAR, Sidery:2013zua, Duverne:2023joq}, the renormalized Jaccard index \citep{jaccardindex}, and the Kullback--Leibler (KL) divergence \citep{KLdivergence}. Skymaps are represented using the \texttt{HEALPix} pixelization scheme \citep{Gorski:2005}. PP plots are computed using full resolution skymaps as provided by \texttt{BAYESTAR}, while we downsample the resolution parameter \texttt{nside} = 256 to compute the Jaccard index and KL divergence, in order to enable fast computation with negligible change in the metric values. This corresponds to an angular pixel size of approximately $0.004$ radians (i.e., $\sim 0.23^\circ$).

\paragraph{Percentile--percentile plots.}
The statistical consistency of the sky-localization posteriors is evaluated using percentile--percentile (PP) plots. 
For each simulated event, we determine the credible region (CR) that encloses the true sky position and construct the corresponding empirical cumulative distribution.  A perfectly calibrated localization yields a diagonal PP plot, while systematic deviations below or above the diagonal indicate overconfident or underconfident posteriors, respectively. 
To quantitatively assess these deviations, we apply a Kolmogorov--Smirnov (KS) test \citep{KStest}, which measures the maximum departure from the expected uniform distribution under the null hypothesis of perfect calibration.

\paragraph{Jaccard Index.} The Jaccard index \( J \) provides a geometric measure of overlap between two sets \( A \) and \( B \) at a given CR, defined as
\begin{equation}
    J = \frac{|A \cap B|}{|A \cup B|},
\end{equation}
where \( A \) and \( B \) denote the \( X\% \) CRs from two-detector and three-detector (HLV) skymaps, respectively. While \( J \) captures the degree of intersection, it is sensitive to differences in contour area. To isolate spatial agreement, we adopt a renormalized form,
\begin{equation}
    J' = \frac{J}{R^{(X)}} = \frac{ |A \cap B| \cdot |A| }{ |A \cup B| \cdot |B| },
\end{equation}
where \( R^{(X)} = |B| / |A| \) is the area ratio. This normalization ensures \( J' = 1 \) for perfectly aligned contours, independently of the sky area reduction. Intermediate values \( 0 < J' < 1 \) indicate partial overlap, while \( J' \rightarrow 0 \) reflects minimal or no spatial agreement between the two localizations. In the following, we focus on the $90\%$ CRs, and thus report values of the renormalized Jaccard index as $J'_{90}$.

\paragraph{Kullback--Leibler Divergence.} While the Jaccard index captures geometric overlap at fixed CRs, it does not reflect the full structure of the posterior distribution. The KL divergence offers a complementary, information-theoretic measure of dissimilarity between probability distributions, accounting for the entire skymap.

Each GW skymap is represented as a discretized posterior probability distribution over the celestial sphere using \texttt{HEALPix}. Given two normalized posteriors $P$ and $Q$, the KL divergence from $Q$ to $P$ is defined as:
\begin{equation}
D_{\mathrm{KL}}(P \parallel Q) = \sum_i P_i \log \left( \frac{P_i}{Q_i} \right),
\end{equation}
where the index $i$ runs over all skymap pixels. This quantity measures the information lost when approximating $P$ with $Q$ and is inherently asymmetric: $D_{\mathrm{KL}}(P \parallel Q) \neq D_{\mathrm{KL}}(Q \parallel P)$ in general.

In our analysis, we compute the KL divergence in both directions:
\begin{equation}
D_{\mathrm{KL}}(\text{2-det} \parallel \text{HLV}), \qquad D_{\mathrm{KL}}(\text{HLV} \parallel \text{2-det}),
\end{equation}
to characterize differences between two-detector and three-detector (HLV) skymaps. The former measures the information lost when approximating the HLV posterior using the two-detector distribution, while the latter quantifies the extent to which the HLV skymap introduces new localization information not supported by the two-detector result.

\section{Injections on Stationary Gaussian Noise}
\label{sec:gaussian_noise}

To establish a baseline for evaluating rapid sky localization, we begin with an idealized simulation set in which all detectors are modeled with Gaussian and stationary noise. This controlled setting isolates the intrinsic performance of \texttt{BAYESTAR} by eliminating complications from non-stationary or glitchy data.

In order to facilitate a direct comparison of which detectors contribute more strongly to the network SNR, we define the per–detector SNR \emph{ratio}
\[
f_{\mathrm{IFO}}
\equiv
\frac{\rho_{\mathrm{IFO}}}{\rho_{\mathrm{net}}}, 
\qquad
\rho_{\mathrm{net}} = \sqrt{\rho_H^2 + \rho_L^2 + \rho_V^2}\, .
\]

Across the simulated population, the median single–detector SNRs are
\(
\rho_H^{\mathrm{med}} = 7.37,\ 
\rho_L^{\mathrm{med}} = 8.58,\ 
\rho_V^{\mathrm{med}} = 3.93
\).
Using the definition above, the corresponding median SNR ratios are
\(
f_H^{\mathrm{med}} = 0.608\ (60.8\%),\ 
f_L^{\mathrm{med}} = 0.698\ (69.8\%),\ 
f_V^{\mathrm{med}} = 0.332\ (33.2\%)
\).
These values summarize the typical relative weight each instrument carries in the Gaussian–stationary baseline, and will be used alongside the absolute SNR to interpret trends in localization performance.

\subsection{Skymap Posterior Calibration}

\begin{figure*}[htbp]
    \centering

    \begin{minipage}[t]{0.48\textwidth}
        \includegraphics[width=\textwidth]{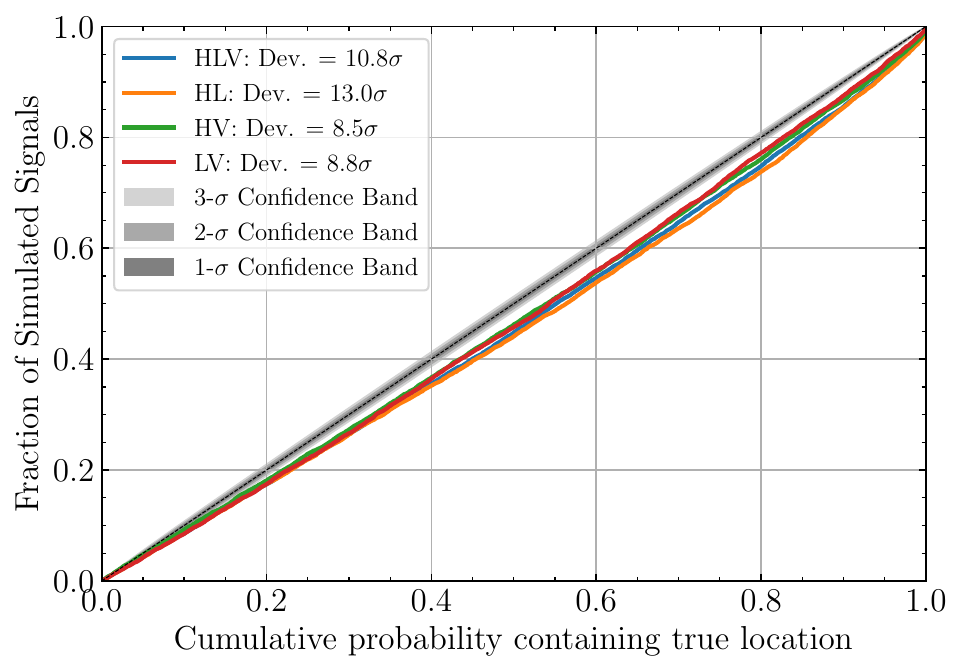}
        \vspace{-0.6cm}
        \begin{center}(a) BBH – Gaussian-stationary\end{center}
    \end{minipage}
    \hfill
    \begin{minipage}[t]{0.48\textwidth}
        \includegraphics[width=\textwidth]{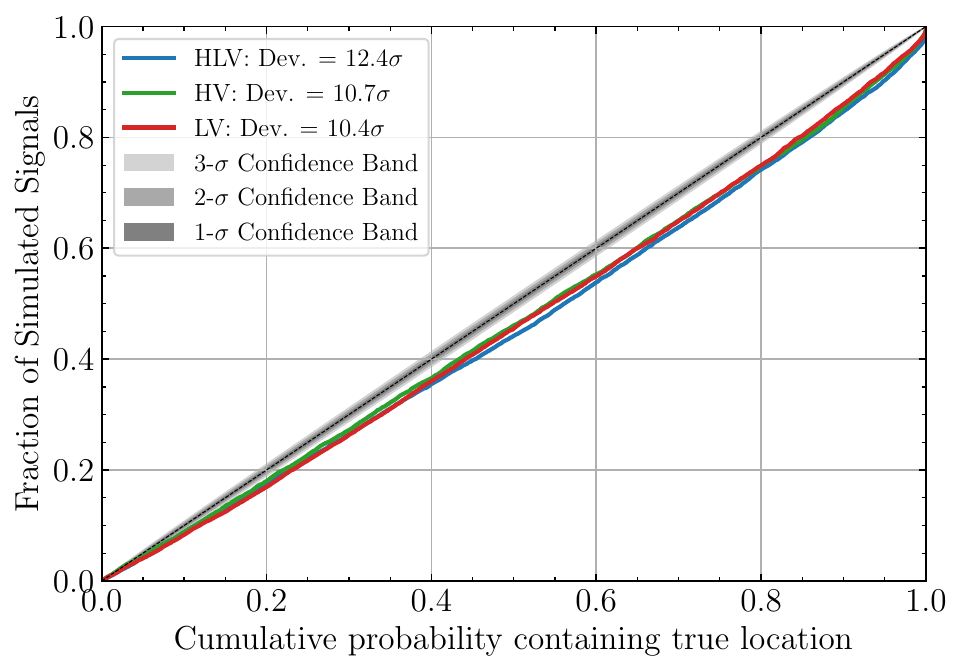}
        \vspace{-0.6cm}
        \begin{center}(b) BBH – Real $h(t)$\end{center}
    \end{minipage}

    \vspace{0.2cm}

    \begin{minipage}[t]{0.48\textwidth}
        \includegraphics[width=\textwidth]{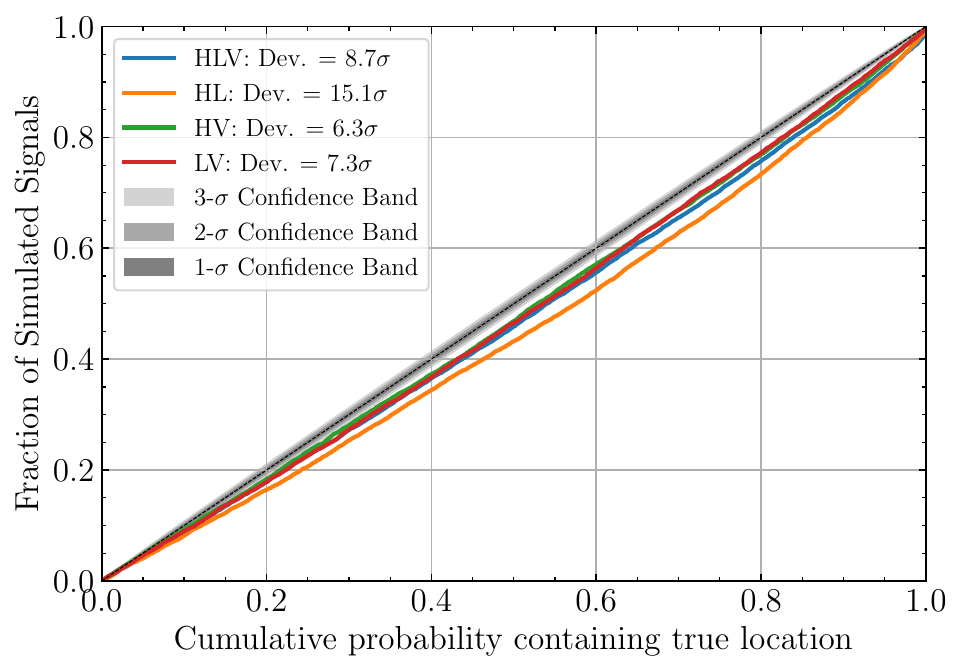}
        \vspace{-0.6cm}
        \begin{center}(c) NSBH – Gaussian-stationary\end{center}
    \end{minipage}
    \hfill
    \begin{minipage}[t]{0.48\textwidth}
        \includegraphics[width=\textwidth]{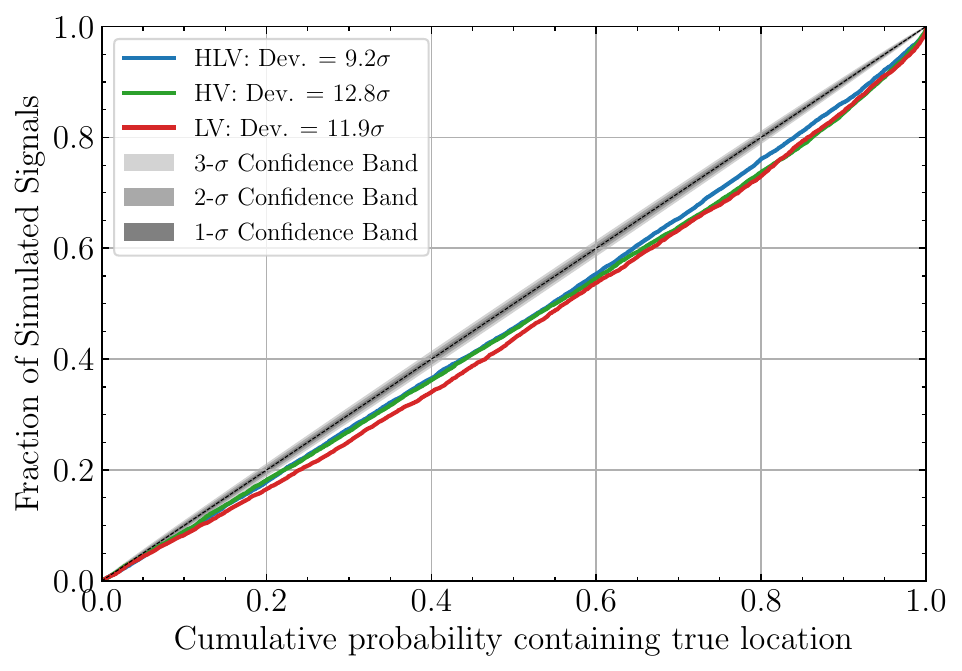}
        \vspace{-0.6cm}
        \begin{center}(d) NSBH – Real $h(t)$\end{center}
    \end{minipage}

    \vspace{0.2cm}

    \begin{minipage}[t]{0.48\textwidth}
        \includegraphics[width=\textwidth]{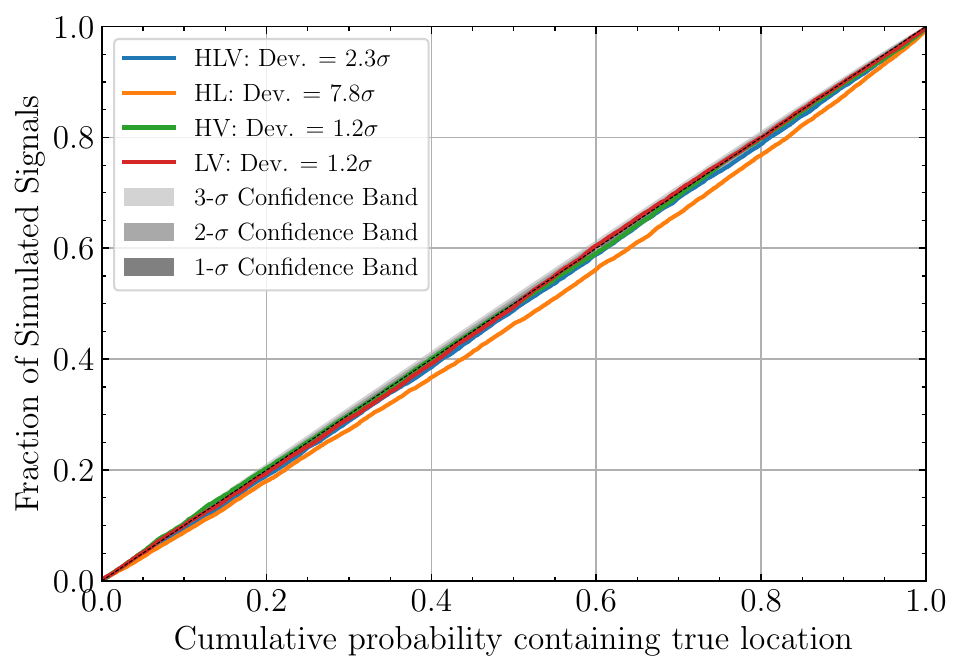}
        \vspace{-0.6cm}
        \begin{center}(e) BNS – Gaussian-stationary\end{center}
    \end{minipage}
    \hfill
    \begin{minipage}[t]{0.48\textwidth}
        \includegraphics[width=\textwidth]{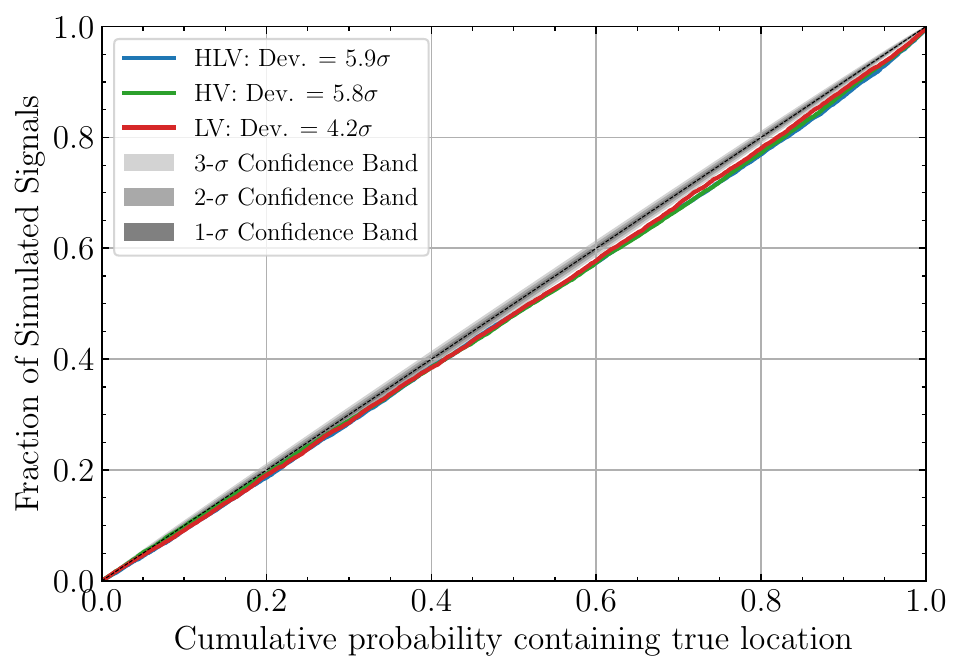}
        \vspace{-0.6cm}
        \begin{center}(f) BNS – Real $h(t)$\end{center}
    \end{minipage}

    \caption{PP plots assessing statistical calibration of sky localization across CBC categories, comparing results between Gaussian-stationary noise in Virgo (left column) and real Virgo strain data (right column). The diagonal line represents ideal calibration. The diagonal line represents ideal calibratio and shaded bands represent 1-, 2-, and 3-sigma confidence intervals.}
    \label{fig:ppplots}
\end{figure*}

\begin{table*}[htbp]
    \centering
    \caption{Kolmogorov--Smirnov test $p$--values and corresponding deviations (in $\sigma$ units) for each detector configuration and source category. Results are shown for simulated signals recovered in Gaussian--stationary noise and on top of real strain data for Virgo.}
    \label{tab:ppplot_ks_summary}
    \begin{tabular}{llcccccc}
        \hline
        Condition & Detectors 
        & BBH $p_{\mathrm{KS}}$ & BBH Dev.
        & NSBH $p_{\mathrm{KS}}$ & NSBH Dev.
        & BNS $p_{\mathrm{KS}}$ & BNS Dev. \\
        \hline\hline
        Gaussian--stationary 
        & HLV & 2.5$\times10^{-27}$ & 10.8$\sigma$
                    & 2.3$\times10^{-18}$ & 8.7$\sigma$
                    & 0.02 & 2.3$\sigma$ \\
        & HL  & 9.0$\times10^{-39}$ & 13.0$\sigma$
                    & 1.5$\times10^{-51}$ & 15.1$\sigma$
                    & 5.5$\times10^{-15}$ & 7.8$\sigma$ \\
        & HV  & 1.3$\times10^{-17}$ & 8.5$\sigma$
                    & 2.5$\times10^{-10}$ & 6.3$\sigma$
                    & 0.24 & 1.2$\sigma$ \\
        & LV  & 1.2$\times10^{-18}$ & 8.8$\sigma$
                    & 4.1$\times10^{-13}$ & 7.2$\sigma$
                    & 0.24 & 1.2$\sigma$ \\
        \hline
        Real $h(t)$ 
        & HLV & 1.7$\times10^{-35}$ & 12.4$\sigma$
                    & 4.3$\times10^{-20}$ & 9.2$\sigma$
                    & 3.3$\times10^{-9}$ & 5.9$\sigma$ \\
        & HV  & 1.3$\times10^{-26}$ & 10.7$\sigma$
                    & 1.2$\times10^{-37}$ & 12.8$\sigma$
                    & 7.5$\times10^{-9}$ & 5.8$\sigma$ \\
        & LV  & 2.5$\times10^{-25}$ & 10.4$\sigma$
                    & 7.3$\times10^{-33}$ & 11.9$\sigma$
                    & 2.6$\times10^{-5}$ & 4.2$\sigma$ \\
        \hline
    \end{tabular}
\end{table*}

Under Gaussian--stationary noise, we evaluate the calibration of the sky-localization posteriors across detector configurations and source categories using PP plots (Fig.~\ref{fig:ppplots}, left). 
The KS statistics, summarized in Table~\ref{tab:ppplot_ks_summary}, quantify the maximum deviation from uniformity expected for a perfectly calibrated posterior distribution. 
The results reveal a clear dependence on the intrinsic source population and network geometry. 
The BBH population exhibits the strongest deviations from the diagonal, followed by NSBH, while BNS injections are generally consistent with ideal statistical calibration. 
Networks including Virgo show improved performance relative to HL baselines. 
For the BNS population, the KS $p$--values for Virgo-inclusive networks (HLV, HV, and LV) all exceed $3\times10^{-3}$, corresponding to deviations smaller than $\sim3\sigma$. 
In contrast, the two-detector H1--L1 configuration yields systematically lower $p$--values across all source classes.

\subsection{Improvement from Adding a Third Detector}

\begin{figure*}[htbp]
    \centering

    \begin{minipage}[t]{0.48\textwidth}
        \includegraphics[width=\textwidth]{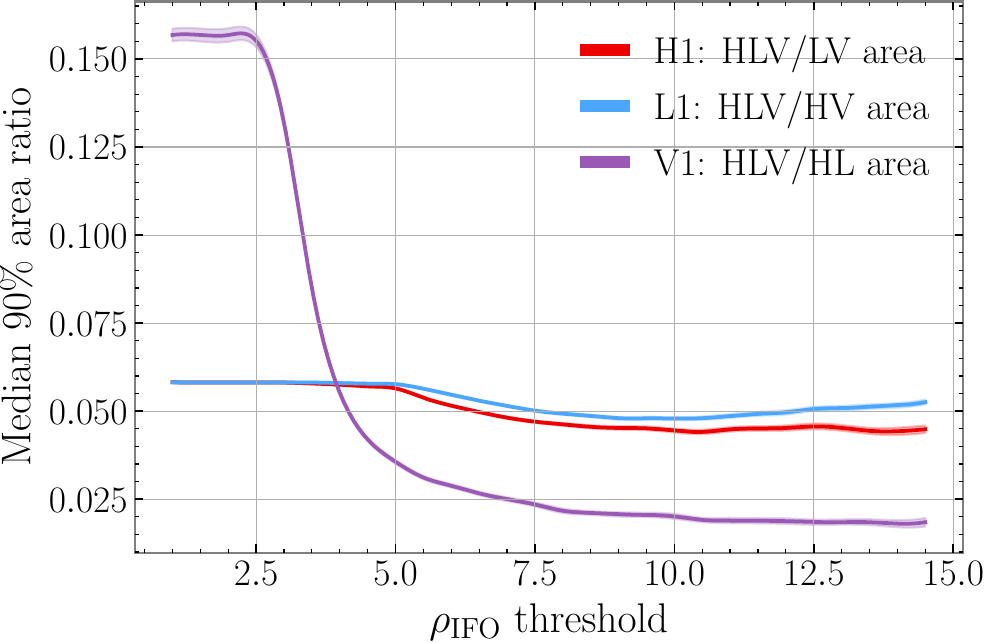}
        \vspace{-0.6cm}
        \begin{center}(a) Median area ratio \( R^{(90)} \) vs. \( \rho_{\mathrm{IFO}} \)\end{center}
    \end{minipage}
    \hfill
    \begin{minipage}[t]{0.48\textwidth}
        \includegraphics[width=\textwidth]{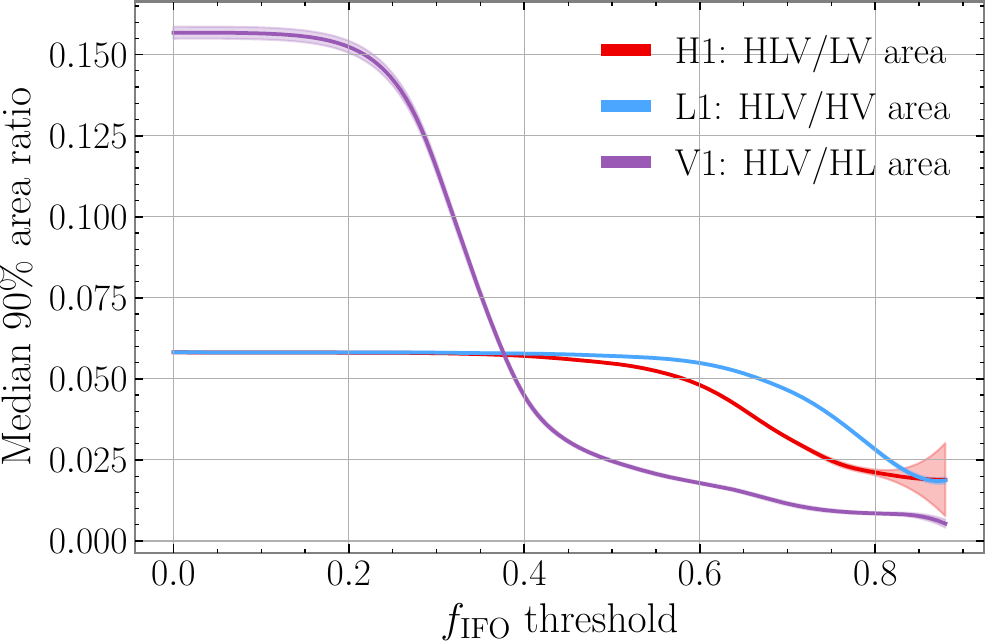}
        \vspace{-0.6cm}
        \begin{center}(b) Median area ratio \( R^{(90)} \) vs. \( f_{\mathrm{IFO}} \)\end{center}
    \end{minipage}

    \vskip\baselineskip

    \caption{Impact of adding a third detector to a 2-detector baseline on the 90\% contour sky area ratio. Each curve corresponds to a specific 2-detector configuration augmented by the third interferometer (H1, L1, or V1). Results include all CBC injections, shaded areas represent 1$\sigma$ error bands.}
    \label{fig:area_reduction_summary}
\end{figure*}

To quantify the gain from extending a two–detector baseline to a three–detector network, we compare the reduction in the 90\% credible sky area \(A^{(90)}\) in each pair (HL, HV, LV) to the corresponding HLV skymap. Fig.~\ref{fig:area_reduction_summary} reports results for the combined CBC population. It shows the median area ratio
\[
R^{(90)} \equiv \frac{A^{(90)}_{\mathrm{HLV}}}{A^{(90)}_{\mathrm{2\text{-}det}}}
\]
as a function of the third–detector SNR threshold \(\rho_{\mathrm{IFO}}\) (left) and of its fractional SNR contribution \(f_{\mathrm{IFO}}\) (right).

\begin{figure*}[htbp]
    \centering
    \begin{minipage}[t]{0.32\textwidth}
        \includegraphics[width=\textwidth]{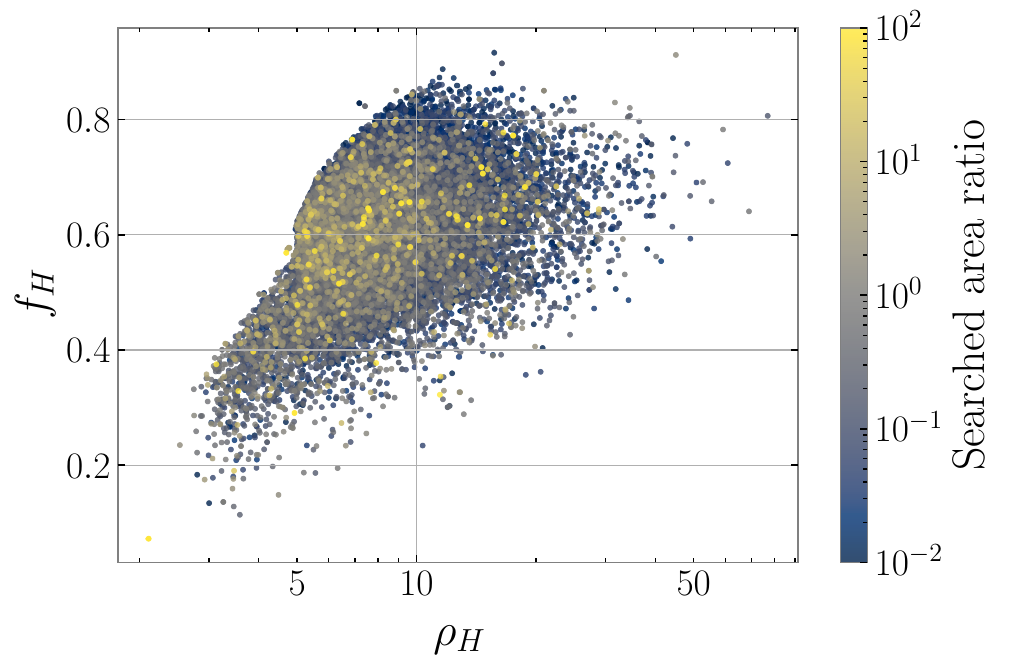}
        \vspace{-0.6cm}
        \begin{center}(a) H1\end{center}
    \end{minipage}
    \hfill
    \begin{minipage}[t]{0.32\textwidth}
        \includegraphics[width=\textwidth]{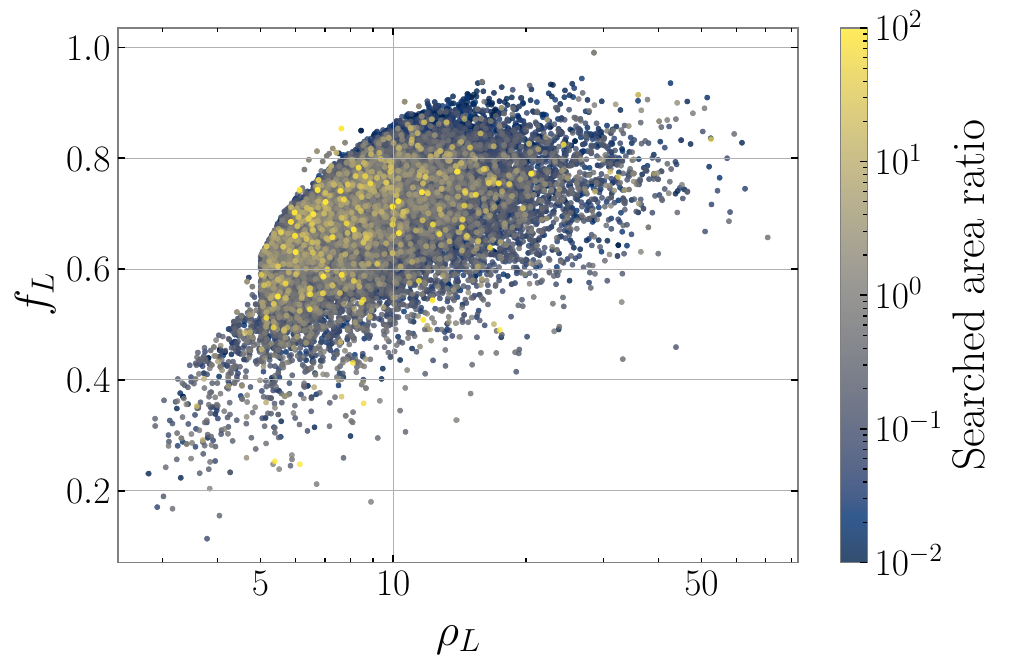}
        \vspace{-0.6cm}
        \begin{center}(b) L1\end{center}
    \end{minipage}
    \hfill
    \begin{minipage}[t]{0.32\textwidth}
        \includegraphics[width=\textwidth]{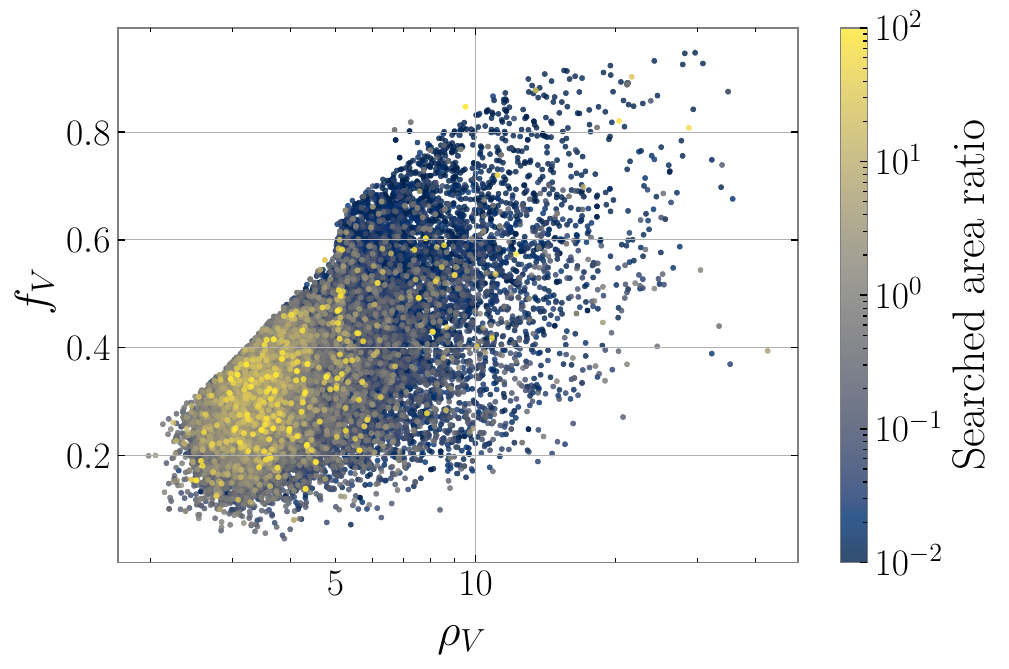}
        \vspace{-0.6cm}
        \begin{center}(c) V1\end{center}
    \end{minipage}
    \caption{SNR ($\rho_{IFO}$) versus SNR fraction ($f_{IFO}$) for each interferometer (H1, L1, V1), color-coded by the searched area ratio. Positive values (yellow) indicate improved localization when adding the third detector; negative values (blue) indicate a degradation.}
    \label{fig:searched_area_diff_snr}
\end{figure*}

For all configurations, adding the third interferometer substantially reduces the localization area. When Virgo is the added detector, the median reduction is \(\sim\!7\times\) and strengthens with increasing \(\rho_V\), reaching \(\sim\!20\times\) for \(\rho_V \gtrsim 4\) (or \(f_V \gtrsim 0.4\)). When H1 or L1 is added, the typical reduction is \(\sim\!20\times\) across thresholds. The apparent plateau for \(\rho_{\mathrm{IFO}}\!\lesssim\!5\) reflects that most injections already exceed this value, so tightening the threshold has little effect on the event set entering the median. The median ratio decreases with \(\rho_{\mathrm{IFO}}\) until \(\rho_{\mathrm{IFO}}\!\sim\!10\), where a mild non–monotonicity appears; this feature is absent when the horizontal axis is expressed in \(f_{\mathrm{IFO}}\), where \(R^{(90)}\) decreases monotonically for \(f_{\mathrm{IFO}}\!\gtrsim\!0.4\). Read together, the two axes indicate that the small non–monotonicity in the \(\rho_{\mathrm{IFO}}\) view arises for injections with more balanced SNR between H1 and L1 at high \(\rho_{\mathrm{IFO}}\).

To further characterize when adding a third detector improves or degrades localization, we compute the ratio of the \emph{searched area} \footnote{We define the searched area as the \emph{smallest} sky area that contains the true location when the sky is scanned in order of decreasing posterior density.} between the three–detector and two–detector skymaps,
\[
R^{(searched)} = A^{(searched)}_{\text{HLV}} / A^{(searched)}_{\text{2-det}}
\]
so that values below 1 indicate improved localization with HLV (smaller offset) and values above 1 indicate degradation. 

Figure~\ref{fig:searched_area_diff_snr} displays, for each instrument, the single–detector SNR versus its SNR fraction, color–coded by $R^{(searched)}$. We observe that the searched area is improved (i.e. lower searched area than in two-detector baseline) in 87.8\% of cases when the added detector is Virgo, and 93.7\% (93.9\%) when it is LIGO Hanford (Livingston). No clear dependence of this improvement rate is observed on $\rho_{\mathrm{H}}$ or $\rho_{\mathrm{L}}$. However, when Virgo has $\rho_{\mathrm{V}}>5$, 96.7\% of skymaps show improved localization, compared to 86\% when $\rho_{\mathrm{V}}<5$.

\subsection{Skymap Overlap: Renormalized Jaccard Index}

\begin{figure*}[htbp]
    \centering
    \begin{minipage}[t]{0.32\textwidth}
        \includegraphics[width=\textwidth]{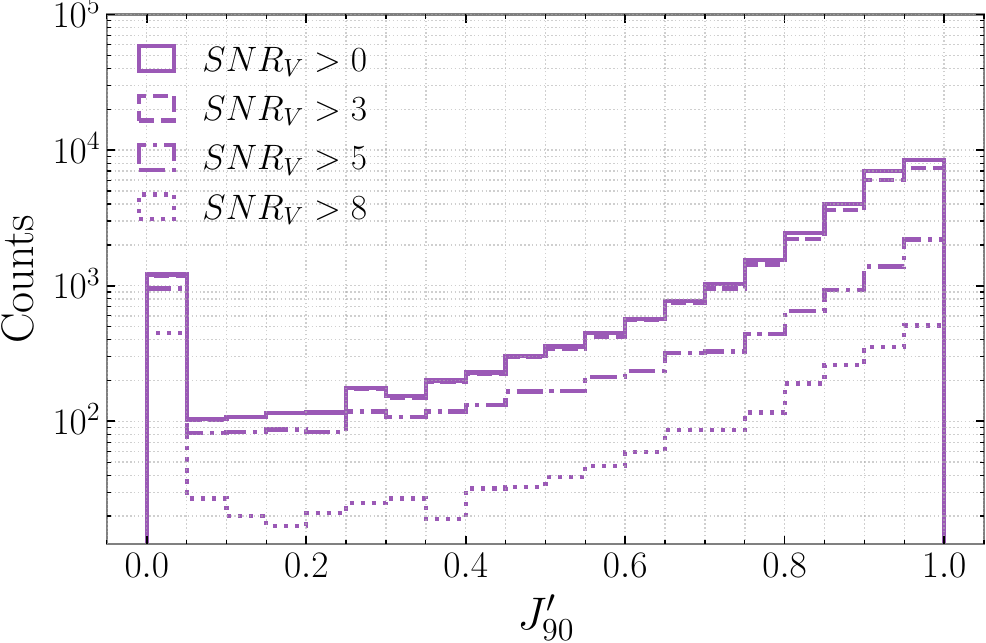}
        \vspace{-0.6cm}
        \begin{center}(a) HL vs. HLV\end{center}
    \end{minipage}
    \hfill
    \begin{minipage}[t]{0.32\textwidth}
        \includegraphics[width=\textwidth]{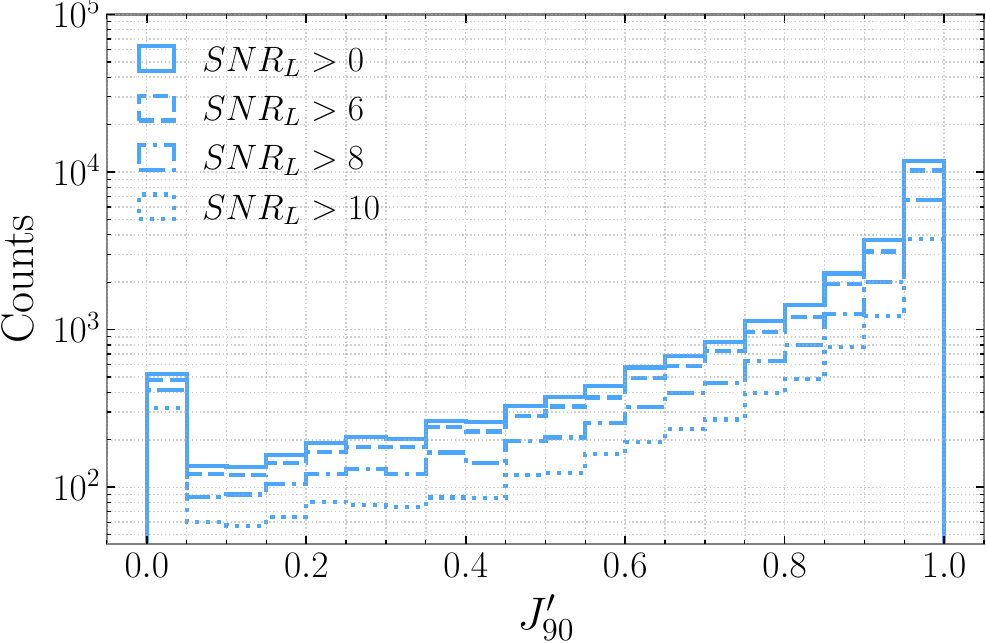}
        \vspace{-0.6cm}
        \begin{center}(b) HV vs. HLV\end{center}
    \end{minipage}
    \hfill
    \begin{minipage}[t]{0.32\textwidth}
        \includegraphics[width=\textwidth]{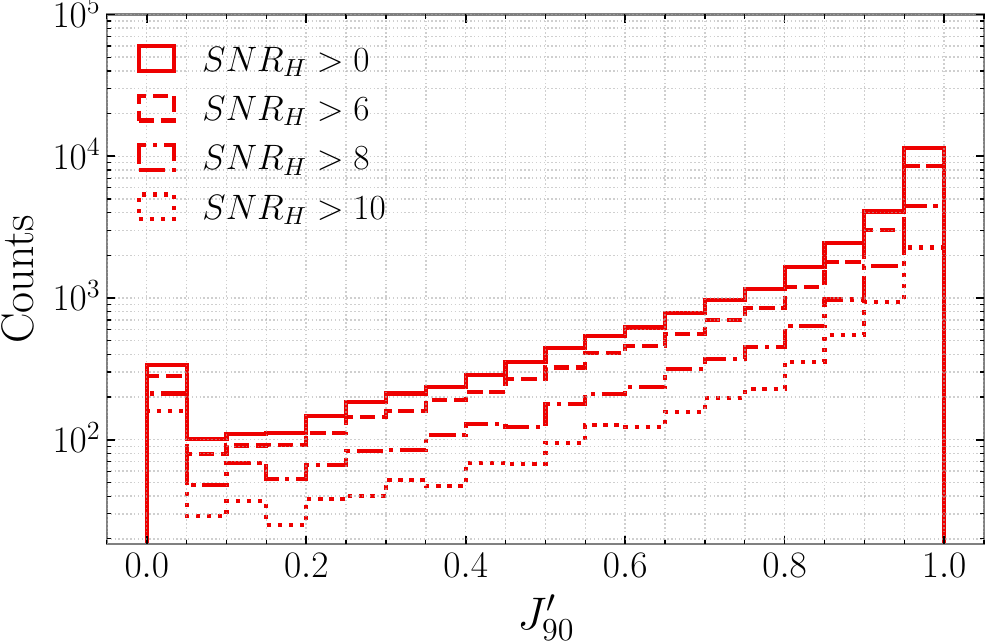}
        \vspace{-0.6cm}
        \begin{center}(c) LV vs. HLV\end{center}
    \end{minipage}
    \caption{Distributions of the renormalized Jaccard index \( J'_{90} \) comparing HLV to HL, HV, and LV skymaps, for all CBC sources. Each panel shows multiple distributions conditioned on increasing SNR thresholds in the third detector.}
    \label{fig:jaccard_distributions_summary}
\end{figure*}

Figure~\ref{fig:jaccard_distributions_summary} shows the distribution of the renormalized Jaccard index at the 90\% CRs, \(J'_{90}\), for the combined CBC population. Each panel compares one two–detector baseline (HL, HV, or LV) against its corresponding HLV skymap, with histograms split by increasing SNR thresholds in the third detector. The index quantifies skymap consistency before and after adding the third instrument: values near unity indicate strong overlap of the 90\% regions, while \(J'_{90}\!\ll\!1\) signals substantial reshaping.

Across configurations, most events cluster near \(J'_{90}\!\approx\!1\), with a secondary accumulation at \(J'_{90}=0\). When H1 or L1 is added, the histogram shape is only weakly dependent on the third–detector SNR. In contrast, when Virgo is added (HL \(\rightarrow\) HLV), higher \(\rho_V\) produces broader \(J'_{90}\) distributions and a more pronounced spike at zero, indicating cases where the 90\% HLV and HL regions become non–overlapping as Virgo’s contribution grows.

Table~\ref{tab:jaccard_summary} summarizes the median \(J'_{90}\) and the fraction of zero–overlap events by source category. The three CBC classes exhibit similar medians and rates, with minor differences. Adding Virgo yields a zero–overlap fraction of \(\sim\!3.5\%\) across categories, whereas adding Hanford or Livingston results in much lower rates (0.5–1.5\%). Thus, LIGO additions typically sharpen the posterior without displacing the peak, while a non–negligible Virgo contribution can, in a small but notable subset, shift the skymap posterior peak and significantly displace the HLV contour. Non–overlapping cases are examined in more detail in \S~\ref{subsec:non_overlapping}.

\begin{table*}
\caption{
Summary of the renormalized Jaccard index analysis on 90\% contours. For each source category (BBH, NSBH, BNS) and baseline (LV, HV, HL), we report the median \( J'_{90} \) and the percentage of cases with no overlap (\( J'_{90} = 0 \)). For the HL configuration, results are shown both for simulations using Gaussian-stationary noise and for those using real Virgo strain data \( h(t) \). The column labeled "\%0" indicates the percentage of cases with \( J'_{90} = 0 \).
}
\label{tab:jaccard_summary}
\centering
\begin{tabular}{lcccccccc}
\hline
 & \multicolumn{2}{c}{HLV vs. LV} & \multicolumn{2}{c}{HLV vs. HV} & \multicolumn{2}{c}{HLV vs. HL (G-S noise)} & \multicolumn{2}{c}{HLV vs. HL (online \( h(t) \))} \\
Category & \( J'_{90} \) & \%0 & \( J'_{90} \) & \%0 & \( J'_{90} \) & \%0 & \( J'_{90} \) & \%0 \\
\hline\hline
BBH  & 0.95 & 0.89 & 0.96 & 1.45 & 0.91 & 3.39 & 0.90 & 2.14 \\
NSBH & 0.95 & 0.95 & 0.96 & 1.50 & 0.90 & 3.78 & 0.89 & 2.26 \\
BNS  & 0.95 & 0.54 & 0.96 & 1.10 & 0.91 & 3.66 & 0.89 & 1.55 \\
\hline
\end{tabular}
\end{table*}

\subsection{Non–overlapping Contours}
\label{subsec:non_overlapping}

\begin{figure}[htbp]
  \centering
  \includegraphics[width=\linewidth]{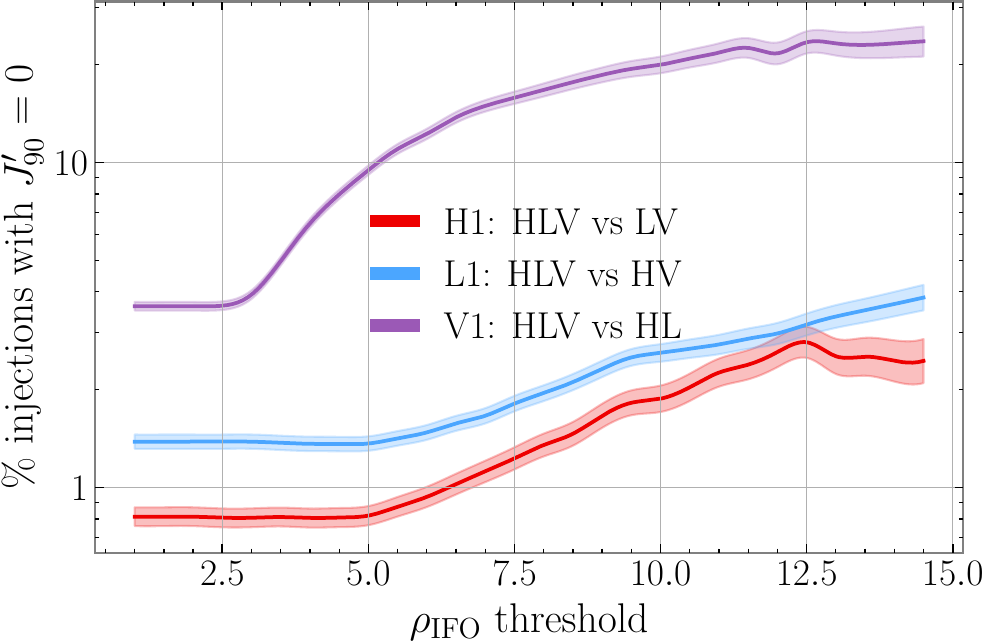}
  
  \vspace{0.6em}

  \includegraphics[width=\linewidth]{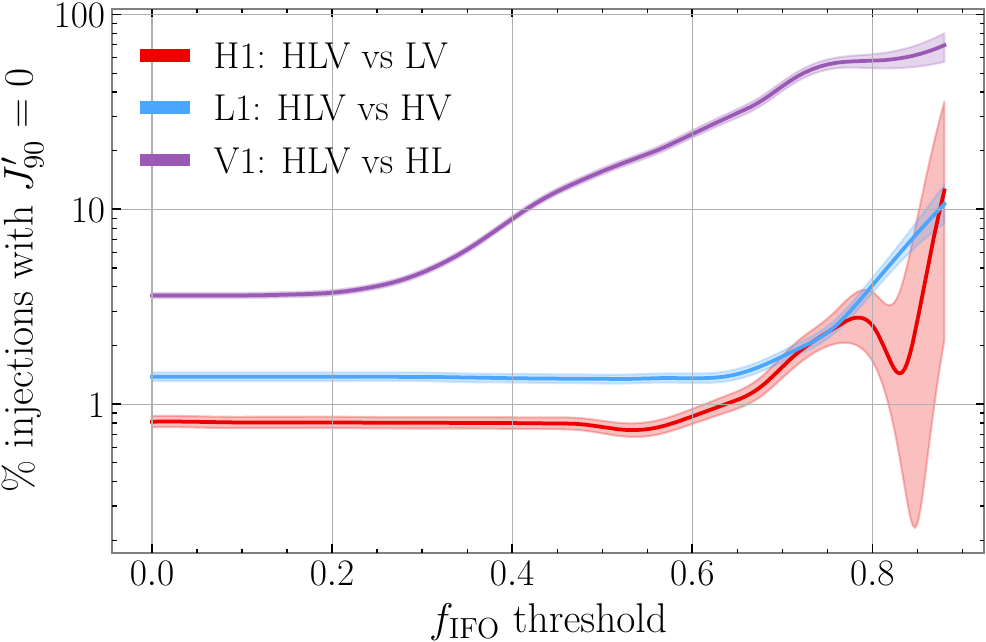}
  \caption{Top: number of events with \( J'_{90} = 0 \) as a function of the third detector’s absolute SNR. Bottom: same quantity as a function of the third detector’s SNR contribution.}
  \label{fig:j0_vs_rho_snrfrac}
\end{figure}

We isolate events in which the 90\% CRs from the two–detector and three–detector reconstructions are disjoint (i.e. \(J'_{90}=0\)). Although uncommon, such cases indicate substantial posterior reweighting when the third instrument is included. We observe that the searched area is improved in 94.3\% of cases when the added detector is Virgo, and 89.6\% (90.4\%) when it is LIGO Hanford (Livingston).  The fraction of non–overlap as a function of the added detector’s SNR \(\rho_{\mathrm{IFO}}\) and fractional contribution \(f_{\mathrm{IFO}}=\rho_{\mathrm{IFO}}/\rho_{\mathrm{net}}\) is shown in Fig.~\ref{fig:j0_vs_rho_snrfrac}. When Virgo is added (HL\(\rightarrow\)HLV), the rate rises steeply with \(f_V\), reaching \(\sim10\%\) for \(\rho_V\gtrsim5\) (or \(f_V\gtrsim0.4\)). When H1 or L1 is added (LV\(\rightarrow\)HLV and HV\(\rightarrow\)HLV), rates remain low, at \(\sim2.5\%\)–\(3\%\) for \(\rho_{\mathrm{IFO}}\gtrsim12\); as a function of \(f_{\mathrm{IFO}}\) they exhibit a broad plateau up to \(\sim0.6\) followed by a mild increase.

\begin{figure}[t]
  \centering
  \includegraphics[width=\linewidth]{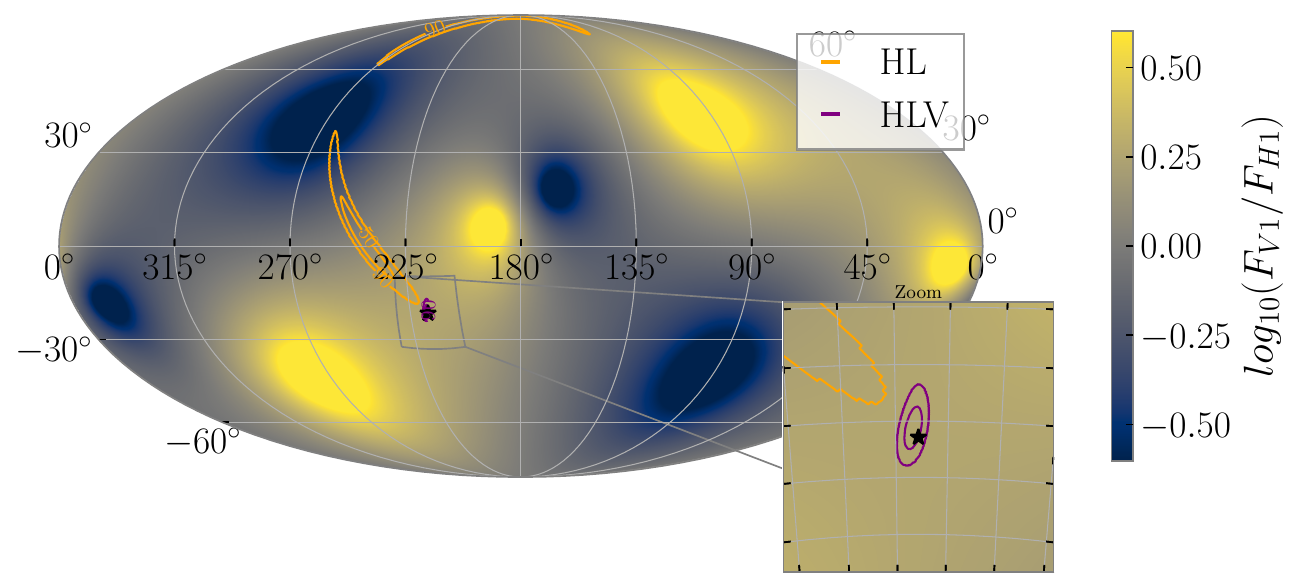}
  \vspace{0.6em}
  \includegraphics[width=\linewidth]{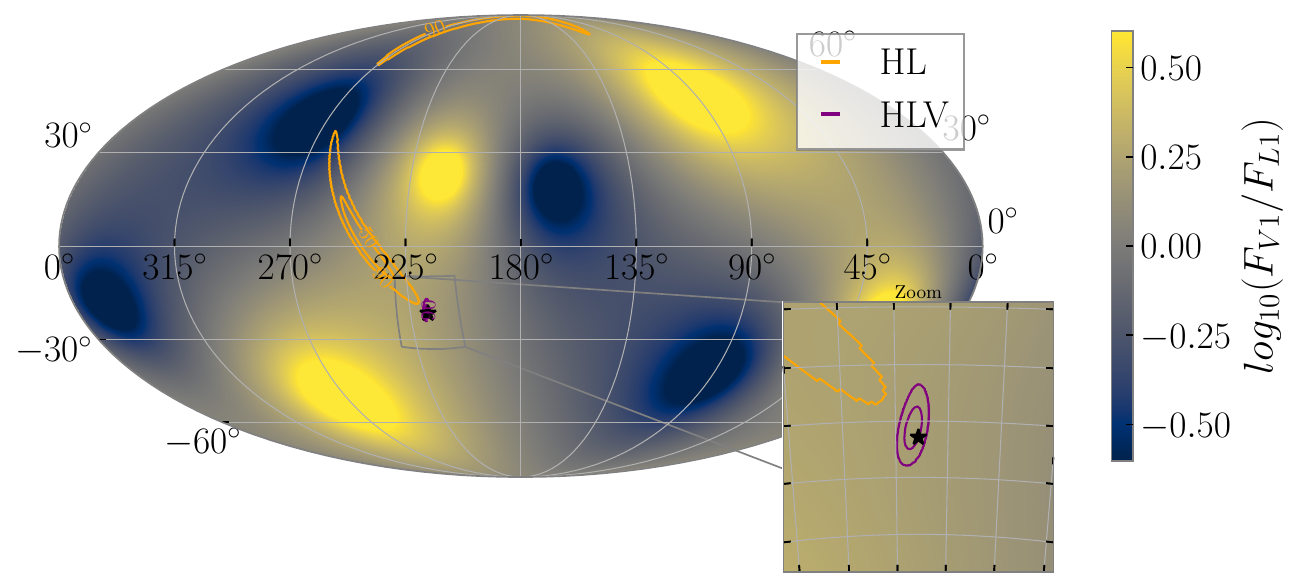}
  \caption{(Top) Antenna-pattern ratio (Virgo/Hanford). (Bottom) Antenna-pattern ratio (Virgo/Livingston). The purple (orange) contour shows the HLV (HL) 90\% and 50\% CR, and the black star marks the injected sky position.}
  \label{fig:nonoverlap_example_skymap}
\end{figure}

\begin{figure}[htbp]
    \centering
    \includegraphics[width=\linewidth]{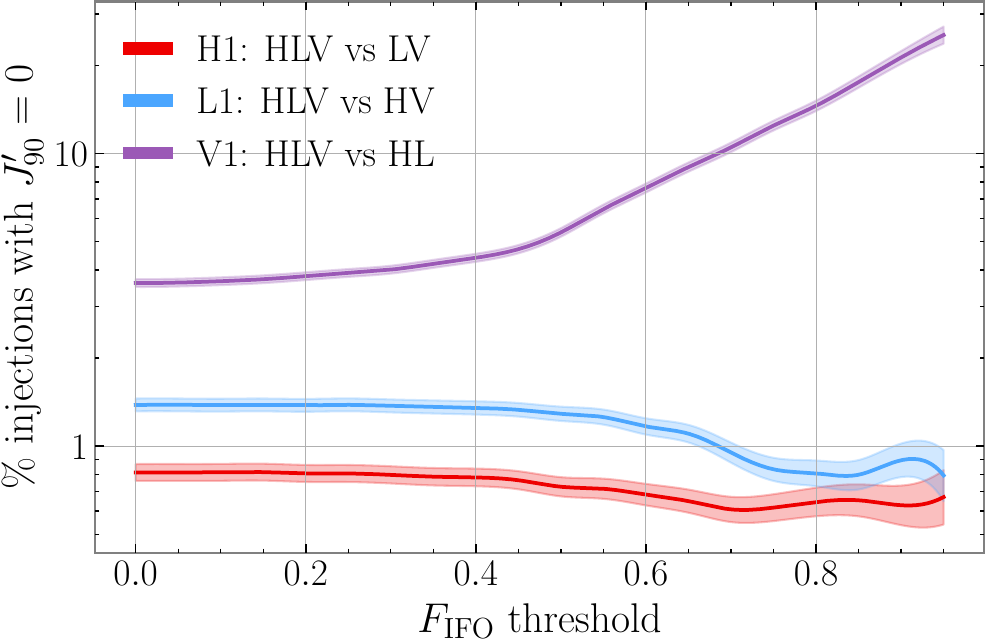}
    \caption{Fraction of events with non-overlapping 90\% CRs versus antenna response threshold of the third detector.}
    \label{fig:jprime_vs_antenna_null_fraction}
\end{figure}

\begin{figure}[h!]
    \centering
    \includegraphics[width=\linewidth]{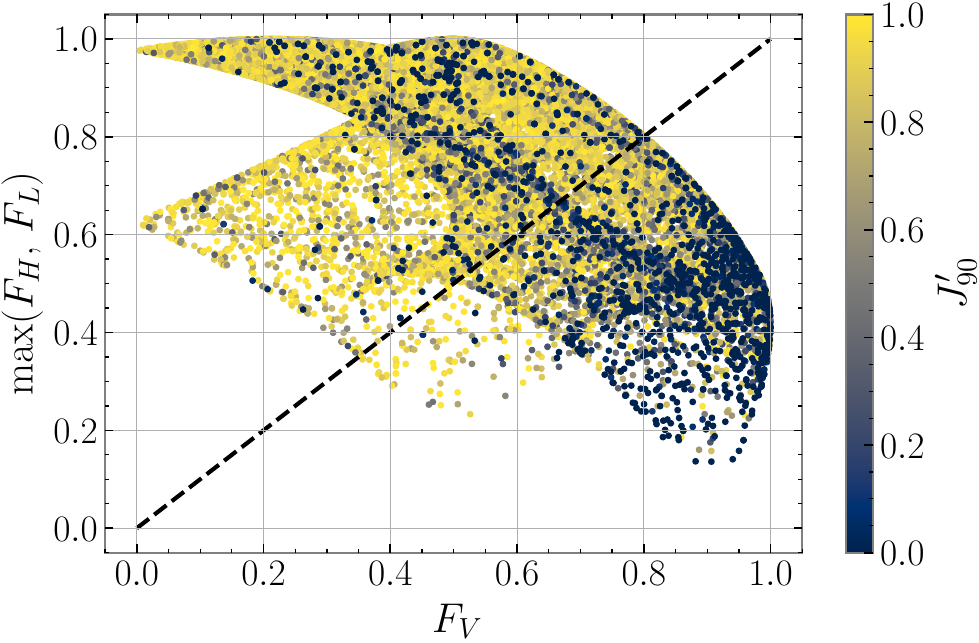}
    \caption{Renormalized Jaccard index \(J'_{90}\) as a function of Virgo’s antenna response \(F_{V1}\) and the maximum LIGO response \(\max\!\big(F_{H1},F_{L1}\big)\).}
    \label{fig:jprime_vs_antenna_fv_fhl}
\end{figure}

For an example BBH injection, Fig.~\ref{fig:nonoverlap_example_skymap} shows maps of the polarization–averaged antenna pattern response ratios \(R_{V1/H1}(\Omega)\equiv F_{V1}(\Omega)/F_{H1}(\Omega)\) (top) and \(R_{V1/L1}(\Omega)\equiv F_{V1}(\Omega)/F_{L1}(\Omega)\) (bottom), where \(F_i(\Omega)\equiv\sqrt{F_{+,i}^2(\Omega)+F_{\times,i}^2(\Omega)}\) is the per–detector antenna factor. The HLV and HL \(90\%\) CRs are overlaid (purple and orange, respectively), and the injected position is marked by a black star. The HL $90\%$ region excludes the true location, whereas including Virgo (HLV) recovers it. Virgo has enhanced directional response toward the source, supplying the additional constraint that shifts posterior support to the correct region. Although the single–detector SNRs in LIGO are high ($\rho_{H}=13.1$, $\rho_{L}=12.2$), Virgo’s moderate contribution ($\rho_{V}=7.6$) provides the independent information needed to resolve the two–detector degeneracy.

Motivated by this example, we examined the dependence on antenna response \(F_{\mathrm{IFO}}\). Figure~\ref{fig:jprime_vs_antenna_null_fraction} shows that the non–overlap fraction increases with \(F_V\), reaching \(\sim10\%\) for \(F_V>0.7\); no clear monotonic trend is observed for H1 or L1 (approximately flat for H1, and for L1 a plateau up to \(F_L\approx0.6\) with a slight decrease thereafter). The complementary scatter in Fig.~\ref{fig:jprime_vs_antenna_fv_fhl} (color–coded \(J'_{90}\) for HLV vs.\ HL) maps these events in the \((F_V,\max(F_H,F_L))\) plane: non–overlap concentrates where \(F_V>\max(F_H,F_L)\), i.e., where Virgo’s antenna response dominates.

These results underscore the central role of the detectors antenna pattern. For sky locations where the H1–L1 baseline has limited directional sensitivity (small \(F_H\) and/or \(F_L\)), adding Virgo with large \(F_V\) -- even at moderate SNR -- provides an independent projection that recenters the posterior on the true position. In this regime, directional coverage, rather than SNR alone, determines the skymap accuracy; a third, differently oriented interferometer is therefore essential to recover accurate localization even when the two–detector SNR is high.

\subsection{Kullback–Leibler Divergence}

\begin{figure*}[htbp]
    \centering
    \begin{minipage}[t]{0.32\textwidth}
        \includegraphics[width=\textwidth]{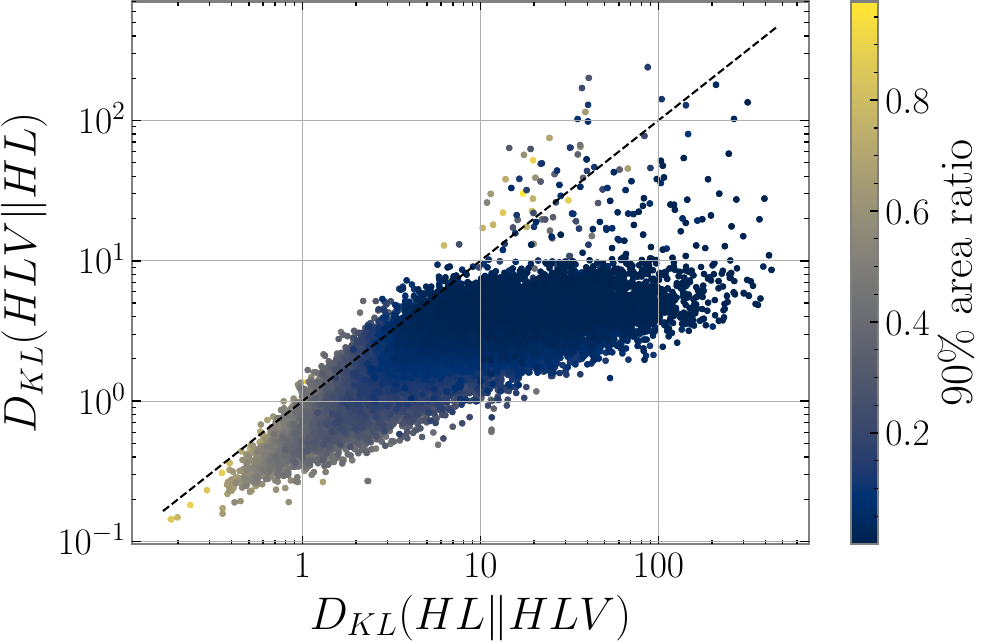}
        \vspace{-0.6cm}
        \begin{center}(a) HL baseline\end{center}
    \end{minipage}
    \hfill
    \begin{minipage}[t]{0.32\textwidth}
        \includegraphics[width=\textwidth]{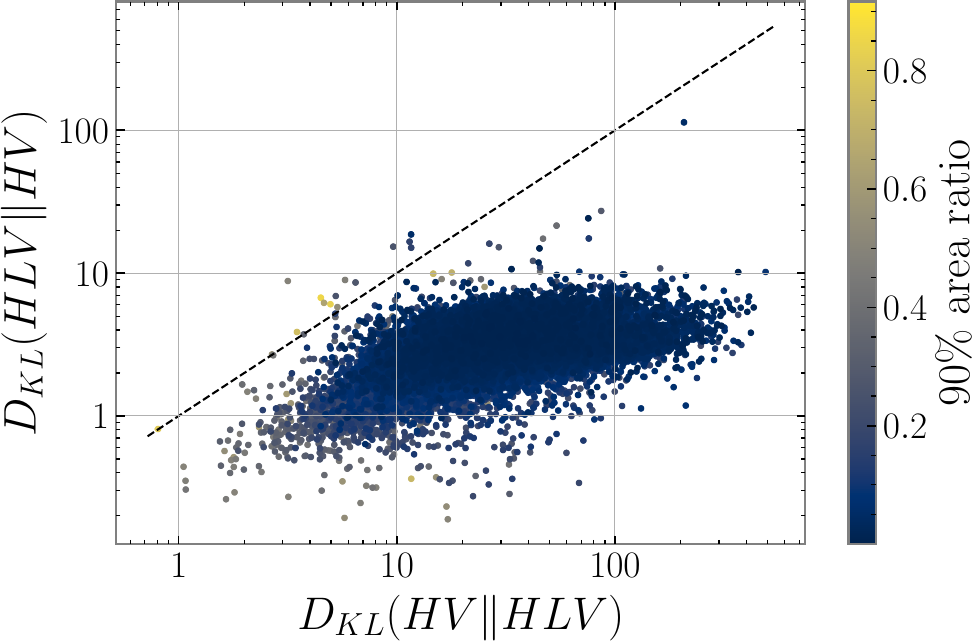}
        \vspace{-0.6cm}
        \begin{center}(b) HV baseline\end{center}
    \end{minipage}
    \hfill
    \begin{minipage}[t]{0.32\textwidth}
        \includegraphics[width=\textwidth]{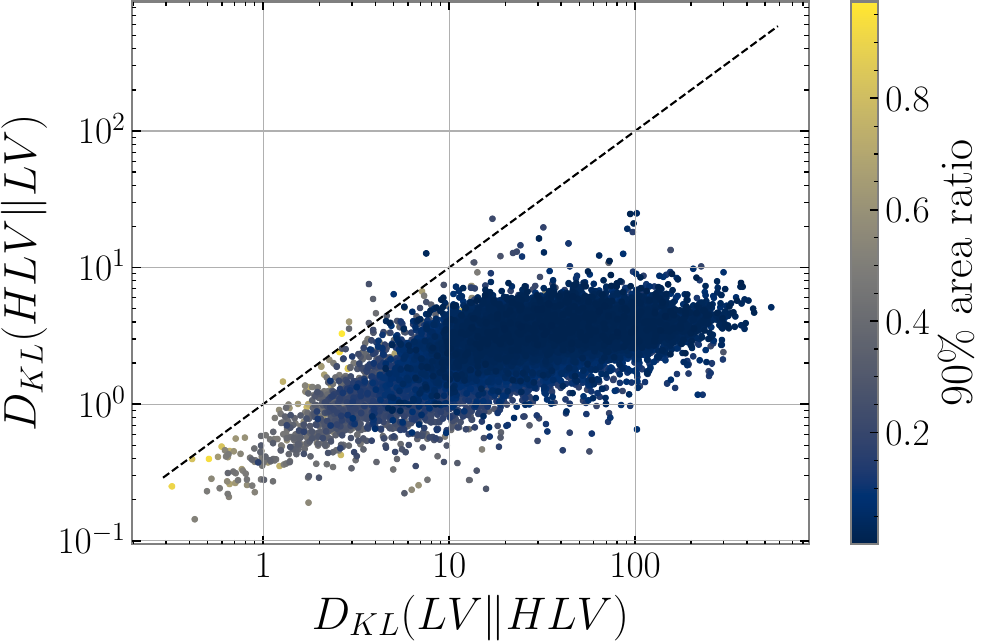}
        \vspace{-0.6cm}
        \begin{center}(c) LV baseline\end{center}
    \end{minipage}
    \caption{Asymmetric KL divergence between two-detector and HLV skymaps, for HL (a), HV (b), and LV (c) baselines. Points are colored by the 90\% area ratio \( R^{(90)} \).}
    \label{fig:kl_scatter}
\end{figure*}

We quantify how posteriors change between two– and three–detector skymaps using the asymmetric Kullback–Leibler divergence, \(D_{\mathrm{KL}}\). Fig.~\ref{fig:kl_scatter} shows, for each two–detector baseline (HL, HV, LV), the pair \(\big(D_{\mathrm{KL}}(\text{2-det}\parallel\text{HLV}),\,D_{\mathrm{KL}}(\text{HLV}\parallel\text{2-det})\big)\), with points color–coded by the 90\% area ratio \(R^{(90)}\).  $D_{\mathrm{KL}}$ increases as the area ratio decreases: events with stronger area reduction (small \(R^{(90)}\)) exhibit systematically larger divergences, reflecting greater information gain from the third detector.

\begin{figure*}[htbp]
    \centering
    \begin{minipage}[t]{0.32\textwidth}
        \includegraphics[width=\textwidth]{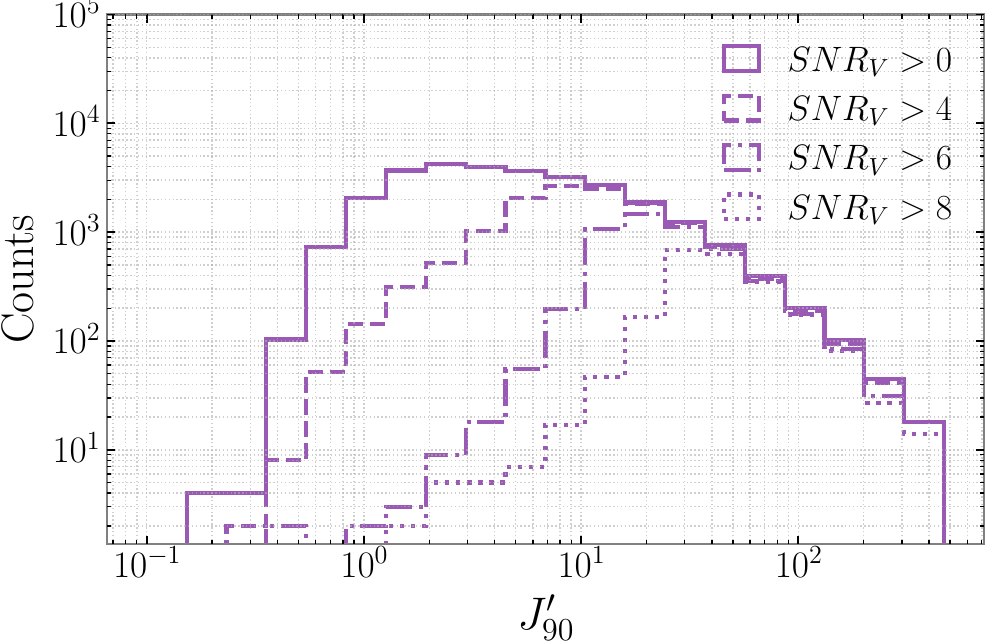}
        \vspace{-0.6cm}
        \begin{center}(a) HL baseline\end{center}
    \end{minipage}
    \hfill
    \begin{minipage}[t]{0.32\textwidth}
        \includegraphics[width=\textwidth]{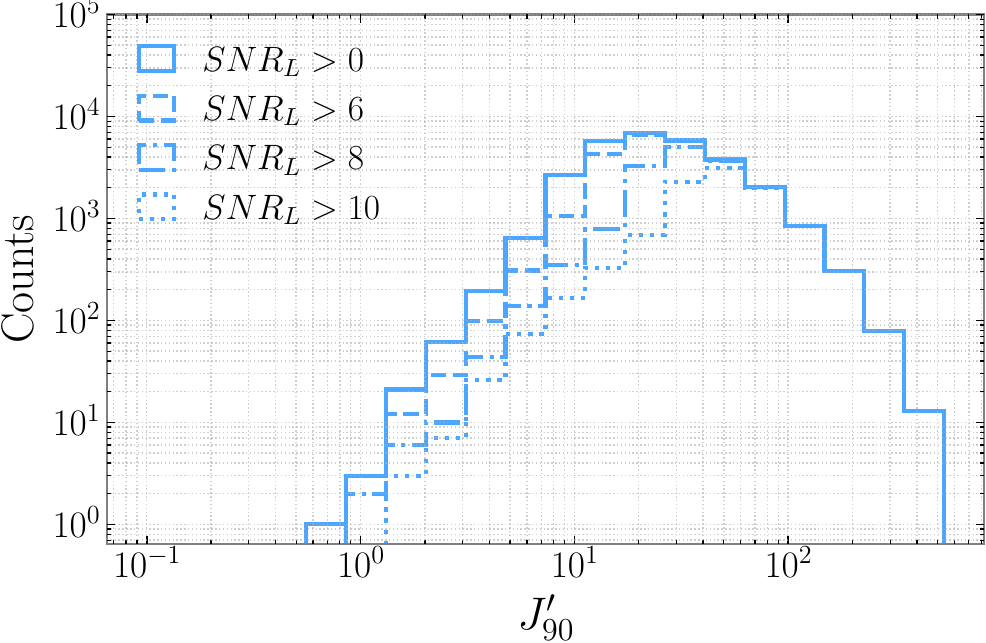}
        \vspace{-0.6cm}
        \begin{center}(b) HV baseline\end{center}
    \end{minipage}
    \hfill
    \begin{minipage}[t]{0.32\textwidth}
        \includegraphics[width=\textwidth]{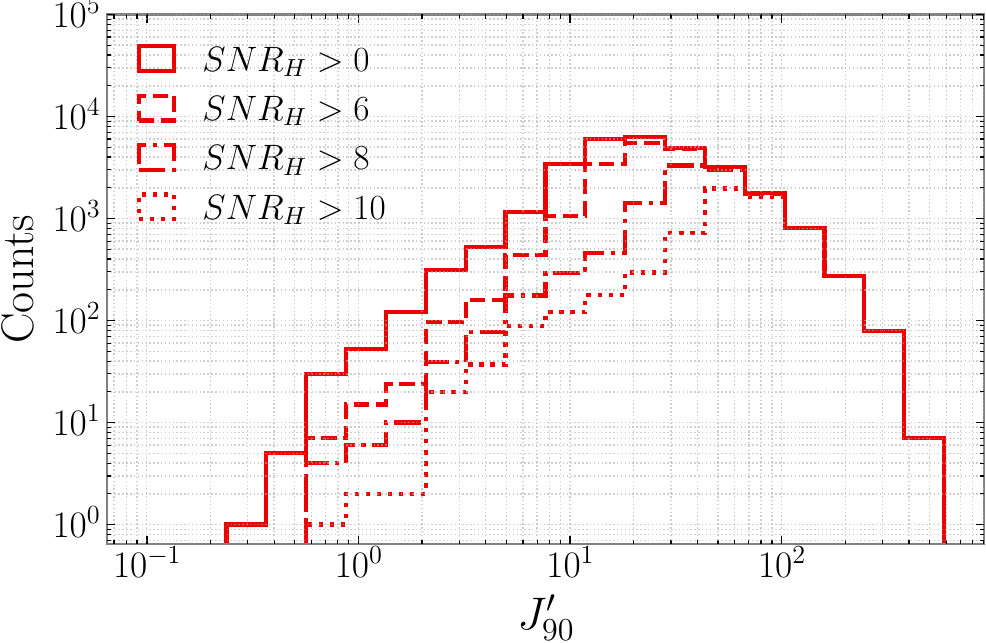}
        \vspace{-0.6cm}
        \begin{center}(c) LV baseline\end{center}
    \end{minipage}
    \caption{Distribution of KL divergence \( D_{\mathrm{KL}}(2\text{-det} \parallel \text{HLV}) \) for different SNR thresholds of the third detector: Virgo for (a), Hanford for (b), and Livingston for (c).}
    \label{fig:kl_histograms}
\end{figure*}

Most injections lie below the diagonal, i.e. \(D_{\mathrm{KL}}(\text{2-det}\parallel\text{HLV}) > D_{\mathrm{KL}}(\text{HLV}\parallel\text{2-det})\), consistent with the three–detector posterior being a sharpened version of the two–detector one at approximately the same location (the broader two–detector map poorly approximates the narrower HLV map, whereas the converse incurs smaller loss). A distinct subset in the HL\(\rightarrow\)HLV comparison appears above the diagonal, indicating posterior displacement: the HLV map concentrates probability in regions of low HL support, in line with the non–overlap behavior discussed in \S~\ref{subsec:non_overlapping}.

Fig.~\ref{fig:kl_histograms} reports \(D_{\mathrm{KL}}(\text{2-det}\parallel\text{HLV})\) distributions under increasing third–detector SNR thresholds. For all baselines the divergence grows with \(\rho_{\mathrm{IFO}}\), showing that the added detector contributes increasingly distinctive information. The effect is most pronounced for HL\(\rightarrow\)HLV: as \(\rho_V\) exceeds \(\sim\!8\), the divergence increases by roughly an order of magnitude, demonstrating that Virgo can substantially restructure the posterior despite its lower sensitivity.

\section{Injections on Real Strain Data in Virgo}
\label{sec:real_noise}

We assess sky localization accuracy using simulated signals injected into Virgo strain data from O3b, while retaining stationary Gaussian noise (built from O3b median ASDs) for Hanford and Livingston. To focus on clean data, we exclude injections with any \texttt{Omicron} transient of \(\mathrm{SNR}>6.5\) within \(\pm1\,\mathrm{s}\) of merger. All diagnostics from the Gaussian benchmark are repeated. In this sample the median Virgo SNR and SNR fraction are \(\rho_V^{\mathrm{med}}=3.64\) and \(f_V^{\mathrm{med}}=0.3044\) (30.44\%), closely matching the Gaussian–stationary case; this supports the representativeness of the O3b median ASDs used in the simulations.

\subsection{Virgo’s Contribution to Localization}
\label{sec:virgo_contribution_clean}

The right panels of Fig.~\ref{fig:ppplots} show the PP–plots for BBH, NSBH, and BNS, and Table~\ref{tab:ppplot_ks_summary} summarizes the corresponding deviations. Relative to the Gaussian benchmark, PP–plot deviations are typically higher with differences across categories and networks. These results highlight the influence of residual low-significance non-Gaussian noise near the coalescence time, as well as the impact of higher-amplitude glitches occurring more than $\pm1$s away from the trigger, which can degrade the calibration of skymap posteriors.

\begin{figure}[htbp]
    \centering
    \includegraphics[width=\linewidth]{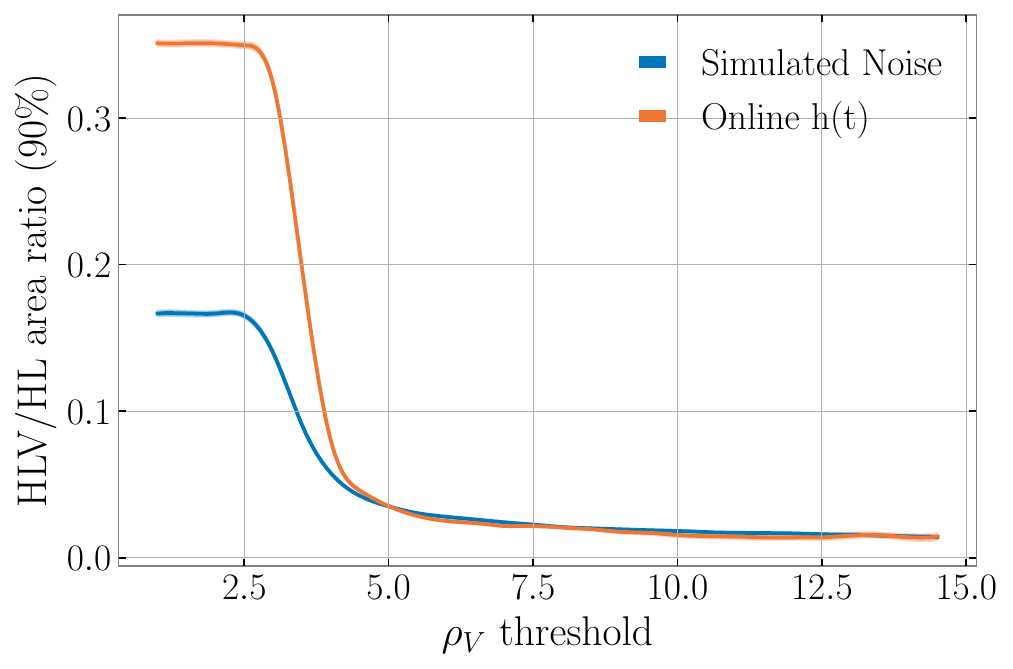}
    \caption{Median area ratio between 90\% HLV and HL CRs as a function of Virgo SNR in real Virgo data compared to the Gaussian-stationary noise case.}
    \label{fig:area_ratio_vs_SNRV}
\end{figure}

\begin{figure}[htbp]
    \centering
    \includegraphics[width=\linewidth]{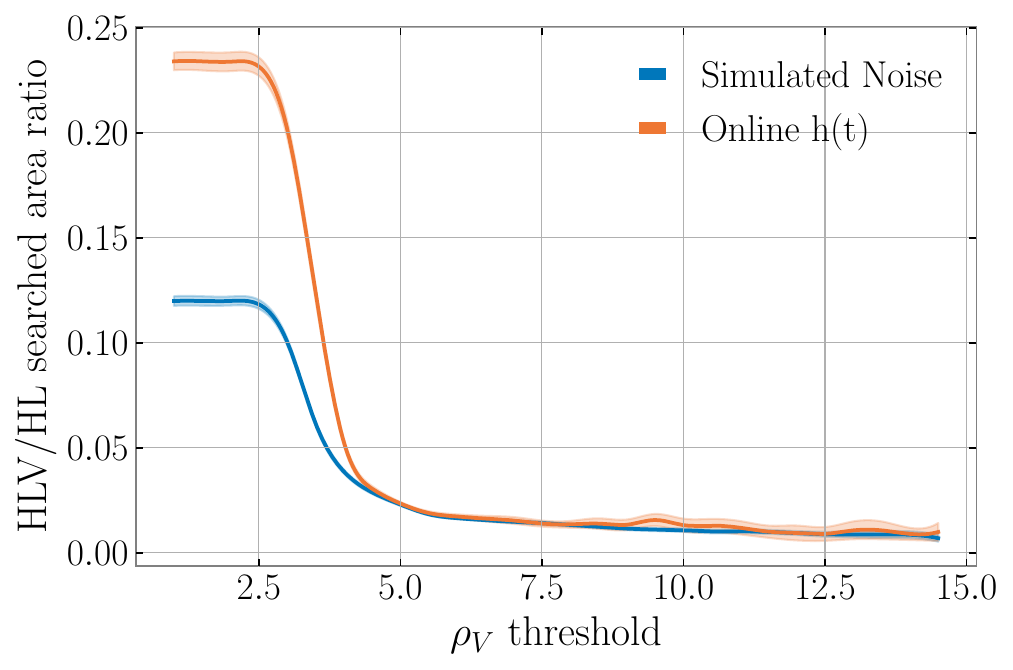}
    \caption{Median searched area ratio between HLV and HL skymaps as a function of Virgo SNR in real Virgo data compared to the Gaussian-stationary noise case.}
    \label{fig:searched_area_ratio_vs_SNRV}
\end{figure}

Fig.~\ref{fig:area_ratio_vs_SNRV} shows the median area ratio \(A_{90}^{\mathrm{HLV}}/A_{90}^{\mathrm{HL}}\) as a function of \(\rho_V\) for simulated (Gaussian–stationary) noise and real O3b Virgo data. Under real strain, Virgo improves localization with a median 90\% area reduction of \(\sim3\times\) (vs.\ \(\sim7\times\) in the Gaussian case). The gap narrows with increasing \(\rho_V\) and becomes small for \(\rho_V\!>\!5\). Analogous behavior is seen for the median searched area ratio in Fig.~\ref{fig:searched_area_ratio_vs_SNRV}. Once Virgo’s SNR exceeds \(\sim5\), the simulated vs. real noise curves converge. For events with \(\rho_V<5\), the fraction exhibiting degraded searched area rises to \(19.8\%\) in real strain, compared with \(14.0\%\) under Gaussian–stationary noise.

\subsection{Skymap Overlap: HLV vs.\ HL}
\label{sec:jaccard_realdata_clean}

We assess the 90\% contour overlap between HLV and HL skymaps using the renormalized Jaccard index \(J'_{90}\). In clean Virgo data, the distribution closely tracks the Gaussian–stationary benchmark: the median is \(J'_{90}\!\approx\!0.89\) across source classes (vs.\ 0.91; Table~\ref{tab:jaccard_summary}), and only \(\sim\!2\%\) of injections yield zero overlap, below the \(\sim\!3.5\%\) observed in Gaussian–stationary noise. 

\begin{figure}[htbp]
    \centering
    \includegraphics[width=\linewidth]{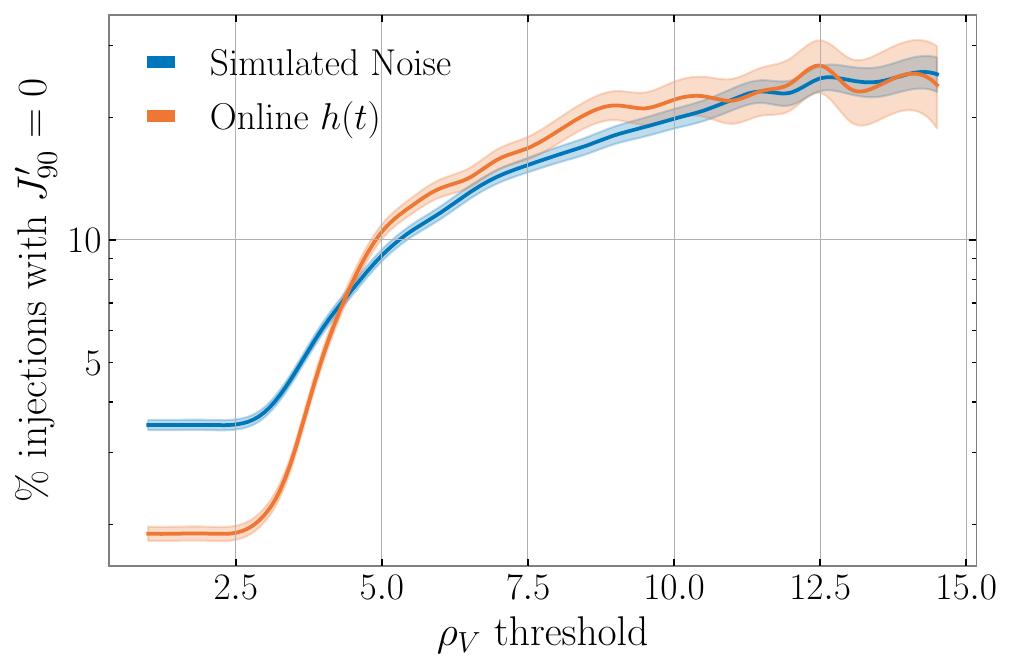}
    \caption{Fraction of injections with completely non-overlapping 90\% CR between HLV and HL skymaps, as a function of Virgo SNR.}
    \label{fig:j0_fraction_real}
\end{figure}

Fig.~\ref{fig:j0_fraction_real} shows the non–overlap fraction as a function of \(\rho_V\). The fraction increases with \(\rho_V\) yet remains below the simulated curve, and for \(\rho_V\!\gtrsim\!4.5\) the real–data and Gaussian–stationary trends become consistent. At low \(\rho_V\), Virgo’s contribution in real strain often reflects residual non–Gaussian features and carries limited astrophysical localization content; the resulting HLV skymaps remain closer to the HL baseline, yielding fewer \(J'_{90}=0\) cases than in simulation. In the Gaussian–stationary case, by contrast, even low–\(\rho_V\) measurements are coherent with the signal and more frequently modify the HL contours; as \(\rho_V\) increases and Virgo’s input becomes signal–dominated, the real–data overlap statistics converge to the Gaussian–stationary expectation.

\section{Interpretation and Implications for Low-Latency Sky Localization}
\label{sec:interpretation}

\paragraph{Skymap posterior overconfidence.}
As shown in \S\ref{sec:gaussian_noise}, the sky localization posteriors exhibit systematic overconfidence across most CBC categories and detector networks, with deviations increasing with the total mass. The largest deviations from ideal calibration are observed for the HL network, while configurations including Virgo show improved consistency. BNS systems, which are the most relevant for electromagnetic follow-up, display the smallest deviations, remaining within $3\sigma$ for the HLV, HV, and LV networks. The origin of the mass dependence remains to be investigated. A plausible explanation is that low-mass binaries generate signals with broader frequency bandwidths, leading to more precise estimates of the inter-detector time delays \citep{Fairhurst:2009tc}, whereas high-mass systems produce shorter, narrower-band signals with larger timing uncertainties. 
If the time-of-flight uncertainties are underestimated by the sky localization algorithm, the resulting sky areas become too small, yielding overconfident credible regions—an effect that becomes more pronounced for high-mass binaries. 
A detailed investigation of this behaviour is left to future work.

\begin{table}[htbp]
    \centering
    \caption{KS test $p$--values and corresponding deviations obtained using 500 injections per source category in the HLV configuration.}
    \label{tab:deviations_duverne_like}
    \begin{tabular}{lcc}
        \hline
        Source category & $p_{\mathrm{KS}}$ & Deviation \\
        \hline\hline
        BNS  & 0.96 & 0.0$\sigma$ \\
        NSBH & 0.05 & 1.9$\sigma$ \\
        BBH  & 0.003 & 2.9$\sigma$ \\
        \hline
    \end{tabular}
\end{table}

In comparison, Fig.~3 of \citet{Duverne:2023joq} presents an independent assessment of posterior calibration based on 500 injections per CBC source category in the HLV configuration, obtained with \texttt{pycbc\_make\_skymap} and $\xi = 1$. 
In that study, the deviations remain below $3\sigma$. 
In \autoref{tab:deviations_duverne_like}, we reproduce the same analysis using a randomly downsampled set of 500 injections for each category. Our results show that the deviations remain within the $3\sigma$ confidence bands, with a maximum deviation of $2.9\sigma$ for BBHs (compared to $1.6\sigma$ in \citealt{Duverne:2023joq}). Table~IV of \citet{Duverne:2023joq} reports the corresponding results for real O3b data, where deviations increase with total mass and reach $3.3\sigma$ for BBHs.

While our study is not designed to assess \texttt{BAYESTAR} performance on real LVK events produced by template-bank searches, an end-to-end mock data challenge (MDC) emulating the LVK O4 low-latency alert infrastructure is presented in \citet{Chaudhary:2023vec}. In this MDC, simulated CBC signals were injected into O3 detector data; events recovered by template-bank searches (\texttt{GstLAL}, \texttt{MBTA}, \texttt{PyCBC}, \texttt{SPIIR}) were then localized and the resulting posteriors evaluated. The PP-plots described in \S4.4 of \citet{Chaudhary:2023vec} \footnote{The associated detector network for each event can be HLV, HL, HV, or LV, depending on observing interferometers. Single-detector triggers are excluded.} show that \texttt{BAYESTAR} credible regions become overconfident at high credibility ($> 3\sigma$), including for BNS signals.

\paragraph{Non-overlapping CRs and antenna response.}
In \texttt{BAYESTAR}, for a trial sky position $\Omega=(\alpha,\delta)$ the likelihood compares the \emph{predicted} inter-site time delays, phases, and relative amplitudes -- obtained from each detector’s antenna patterns $(F_{+,i},F_{\times,i})$, orientation, and location -- to the values measured by the matched-filter search, marginalizing over extrinsic parameters \citep{Singer2016_BAYESTAR}. With two nearly coaligned sites (e.g., HL), the antenna responses are \emph{strongly correlated} -- their $(F_{+},F_{\times})$ maps are nearly proportional over much of the sky -- so signal properties provide limited additional discrimination and many sky locations satisfy the HL constraints, yielding extended (sometimes multi-modal) CRs. Adding a third, differently oriented baseline with \emph{significant} antenna response (e.g., Virgo) supplies an independent projection that eliminates most HL-admissible solutions and concentrates support where all three detectors are jointly consistent. When Virgo’s constraint is sufficiently informative, the HLV posterior maximum can shift outside the HL most probable region, yielding disjoint $90\%$ CRs (non-overlap; $J'_{90}=0$). Similar, but rarer, cases arise when comparing HLV to HV or LV (see \S\ref{subsec:non_overlapping}). In such situations, omitting a geometrically complementary detector can center the skymap on a region that excludes the true source location, leading to mispointed follow-up and potential non-coverage of electromagnetic or neutrino counterparts in multi-messenger searches.

\paragraph{Virgo’s contribution at low SNR.}
For $\rho_V<5$, adding Virgo \emph{degrades} the searched area in $14.0\%$ of injections with stationary Gaussian noise, this rate grows to $19.8\%$ when using real O3b Virgo strain. In such SNR regime, the GW signal is close to the background noise level, and the peak-SNR time associated with the astrophysical event in LIGO can coincide with noise fluctuations in Virgo, producing incoherent timing/phase information and displacing the HLV posterior maximum relative to the well-localized HL. The higher fraction in real data is consistent with residual non-Gaussian features increasing the noise rate.

\paragraph{High-SNR LIGO behavior.}
When both LIGO sites have high and comparable SNRs (e.g., $\rho_H,\rho_L\gtrsim12.5$ with $|\rho_H-\rho_L|/\rho_{\rm net}\ll1$), the two-detector baseline is already well-localized. Adding the other LIGO detector then yields diminishing returns in sky-area reduction: the median $A^{90}_{\rm HLV}/A^{90}_{\rm 2\text{-det}}$ plateaus and can show mild non-monotonicity as the added-detector SNR increases. In this regime, the dominant effect of the added LIGO is to reposition the posterior maximum using the additional timing/phase constraint, rather than to further contract the CR. In these regime, we observe that the searched area decreases even as the area reduction saturates.
\section{Sky Localization of Public Alerts from O3}
\label{sec:o3_alerts}

During the third observing run (O3), the LVK Collaboration issued 80 public alerts for candidate gravitational–wave events. Following offline data–quality vetting, 24 candidates were retracted due to instrumental or environmental artifacts, yielding a final sample of 56 alerts. We analyze these events with the same methodology as for the simulations.

\subsection{Sky–Area Improvement}
\label{sec:skyloc_improvement_O3}

\begin{figure}[htbp]
    \centering
    \begin{minipage}[t]{\linewidth}
        \includegraphics[width=\textwidth]{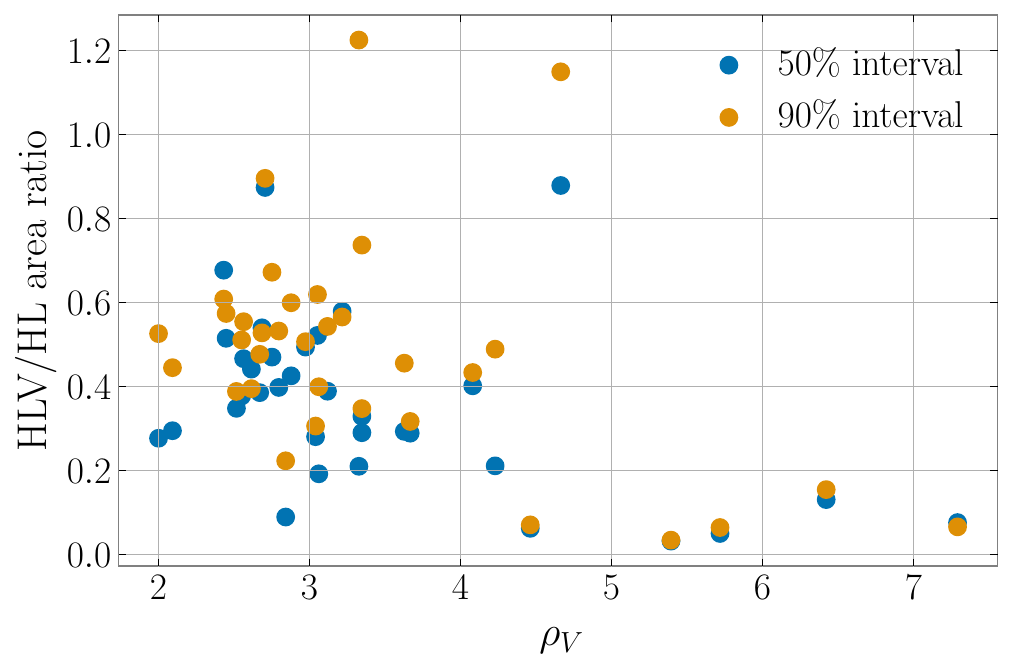}
        \vspace{-0.6cm}
        \begin{center}(a) HL vs. HLV\end{center}
    \end{minipage}
    \vspace{0.6em}
    \begin{minipage}[t]{\linewidth}
        \includegraphics[width=\textwidth]{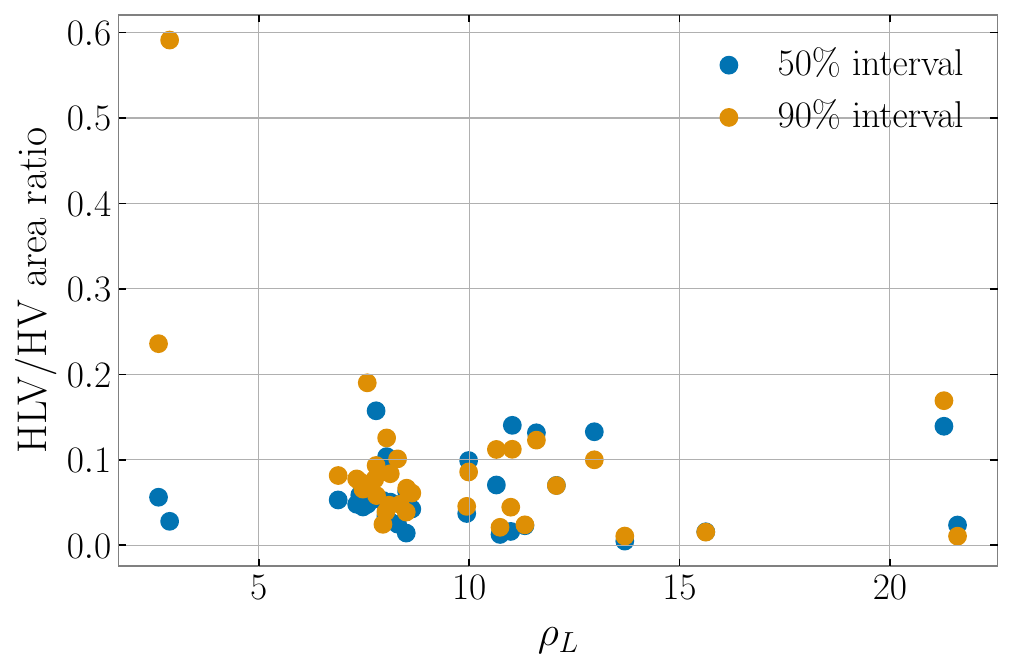}
        \vspace{-0.6cm}
        \begin{center}(b) HV vs. HLV\end{center}
    \end{minipage}
    \vspace{0.6em}
    \begin{minipage}[t]{\linewidth}
        \includegraphics[width=\textwidth]{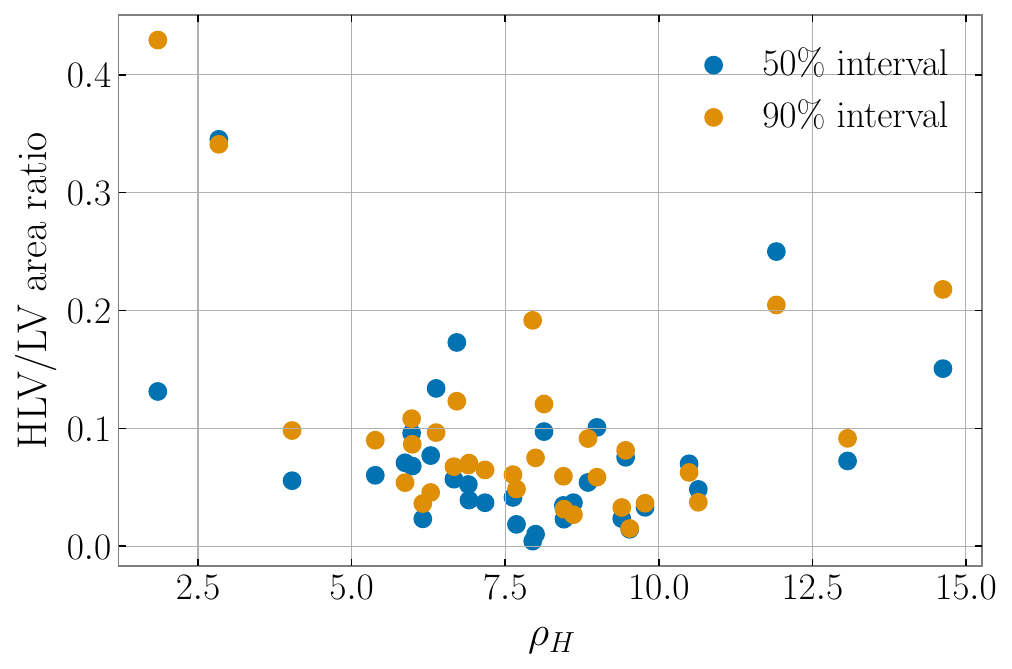}
        \vspace{-0.6cm}
        \begin{center}(c) LV vs. HLV\end{center}
    \end{minipage}
    \caption{Sky area ratio between HLV and two-detector baselines (HL, HV, LV) as a function of the third interferometer's SNR, at 90\% (blue) and 50\% (orange) CRs.}
    \label{fig:skyloc_O3_area_ratio}
\end{figure}

We focus on the 34 alerts observed with the full H1–L1–V1 network. For these events the median single–detector SNRs are \(\rho_V=2.97\), \(\rho_H=7.95\), and \(\rho_L=8.49\). Fig.~\ref{fig:skyloc_O3_area_ratio} reports the ratio \(R^{(p)} \equiv A^{(p)}_{\mathrm{HLV}}/A^{(p)}_{\mathrm{2\text{-}det}}\) at \(p=90\%\) and \(50\%\), comparing HLV to each two–detector baseline. Median ratios are most modest when adding Virgo to HL (HLV vs.\ HL: \(R^{(90)}\!=\!0.50\), \(R^{(50)}\!=\!0.64\)), and become striking when adding a LIGO detector to a Virgo pair (HLV vs.\ HV: \(R^{(90)}\!=\!0.072\), \(R^{(50)}\!=\!0.049\); HLV vs.\ LV: \(R^{(90)}\!=\!0.070\), \(R^{(50)}\!=\!0.056\)), corresponding to reductions by factors of \(\sim\!14\)–20. 

In all comparisons the area ratio decreases with increasing SNR of the added detector, mirroring the trends found in the injection study and reflecting the enhanced triangulation constraints provided by the third interferometer. While most events benefit from the addition of Virgo, we identify two candidates -- \texttt{S190510g} and \texttt{S190718y} -- for
which the 90\% sky area is increased in the HLV configuration.

\subsection{Posterior Consistency}
\label{sec:skyloc_consistency_O3}

\begin{figure}[htbp]
    \centering
    \begin{minipage}[t]{\linewidth}
        \includegraphics[width=\textwidth]{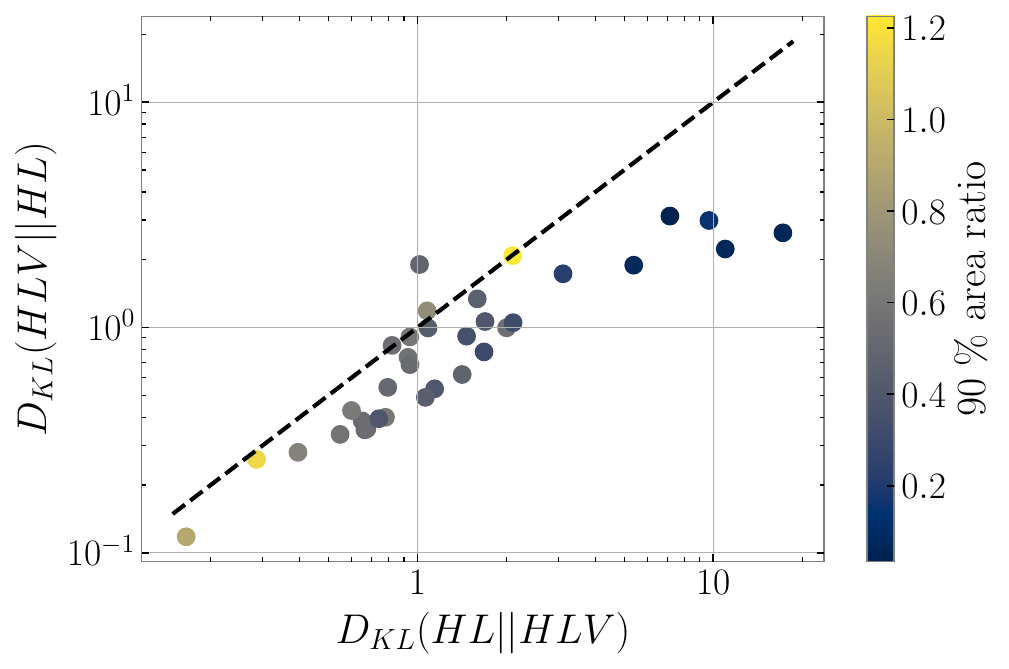}
        \vspace{-0.6cm}
        \begin{center}(a) HL vs. HLV\end{center}
    \end{minipage}
    \vspace{0.6em}
    \begin{minipage}[t]{\linewidth}
        \includegraphics[width=\textwidth]{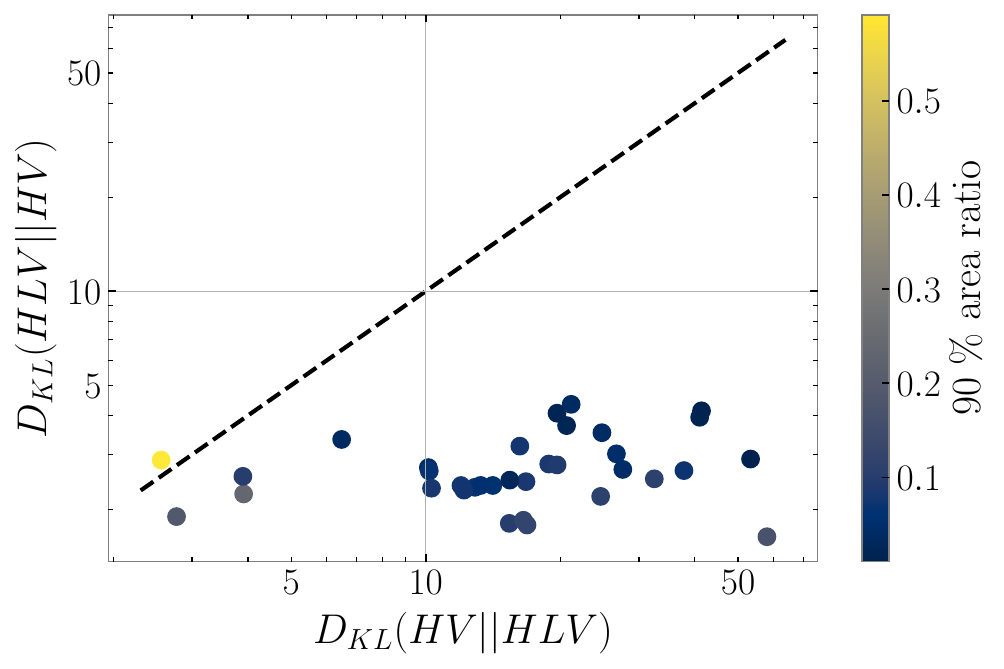}
        \vspace{-0.6cm}
        \begin{center}(b) HV vs. HLV\end{center}
    \end{minipage}
    \vspace{0.6em}
    \begin{minipage}[t]{\linewidth}
        \includegraphics[width=\textwidth]{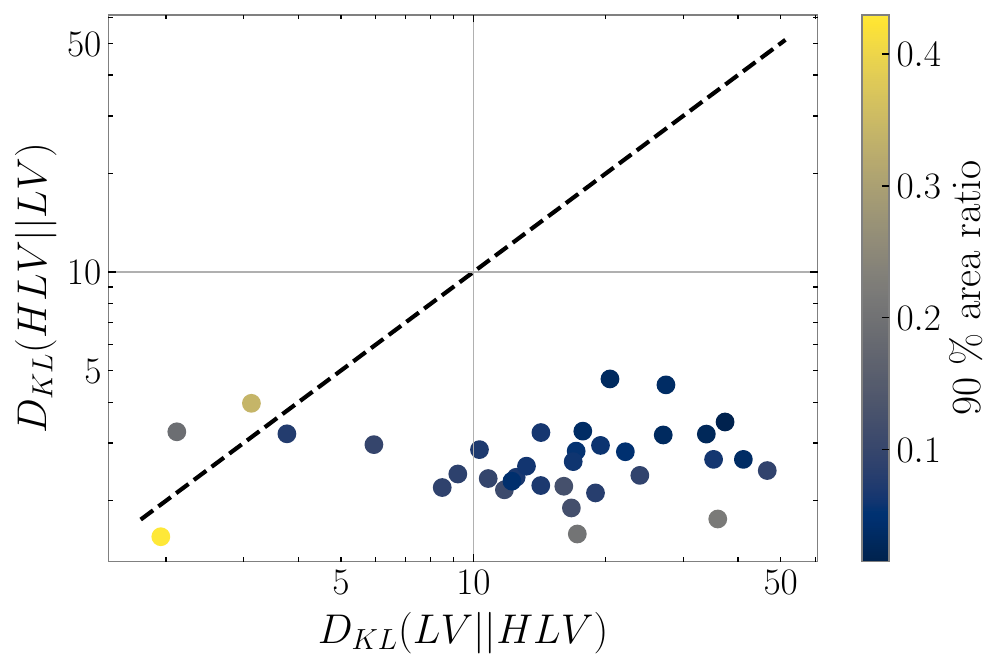}
        \vspace{-0.6cm}
        \begin{center}(c) LV vs. HLV\end{center}
    \end{minipage}
    \caption{KL divergence values comparing each two-detector baseline (HL, HV, LV) with HLV for O3 alerts.}
    \label{fig:skyloc_O3_kl}
\end{figure}

We quantify changes in the skymap between baselines using the bidirectional Kullback–Leibler divergence and the renormalized Jaccard index on the 90\% CRs. Fig.~\ref{fig:skyloc_O3_kl} plots \(\big[D_{\mathrm{KL}}(\text{2-det}\parallel\text{HLV}),\,D_{\mathrm{KL}}(\text{HLV}\parallel\text{2-det})\big]\) for the HL, HV, and LV baselines. In the HL\(\rightarrow\)HLV comparison, most points lie close to the diagonal, indicating relatively symmetric divergence: Virgo both sharpens the posterior and shifts its peak, i.e., it contributes complementary directional information to an already well–localized LIGO pair. By contrast, the HV\(\rightarrow\)HLV and LV\(\rightarrow\)HLV clouds concentrate well below the diagonal, with \(D_{\mathrm{KL}}(\text{HV}\parallel\text{HLV}) \gg D_{\mathrm{KL}}(\text{HLV}\parallel\text{HV})\) (and similarly for LV), consistent with substantial sharpening when a LIGO detector is added but limited displacement of the peak. Events with minimal sky–area change (area ratios near unity) exhibit low divergence in both directions, whereas a small subset above the diagonal -- e.g., \texttt{S200316bj}, \texttt{S200208q}, \texttt{S200213t} in HLV vs.\ HL; \texttt{S190718y} in HLV vs.\ HV; and \texttt{S190718y}, \texttt{S190510g} in HLV vs.\ LV -- shows non–negligible peak shifts when the third detector is included.

\begin{figure}[htbp]
    \centering
    \includegraphics[width=\linewidth]{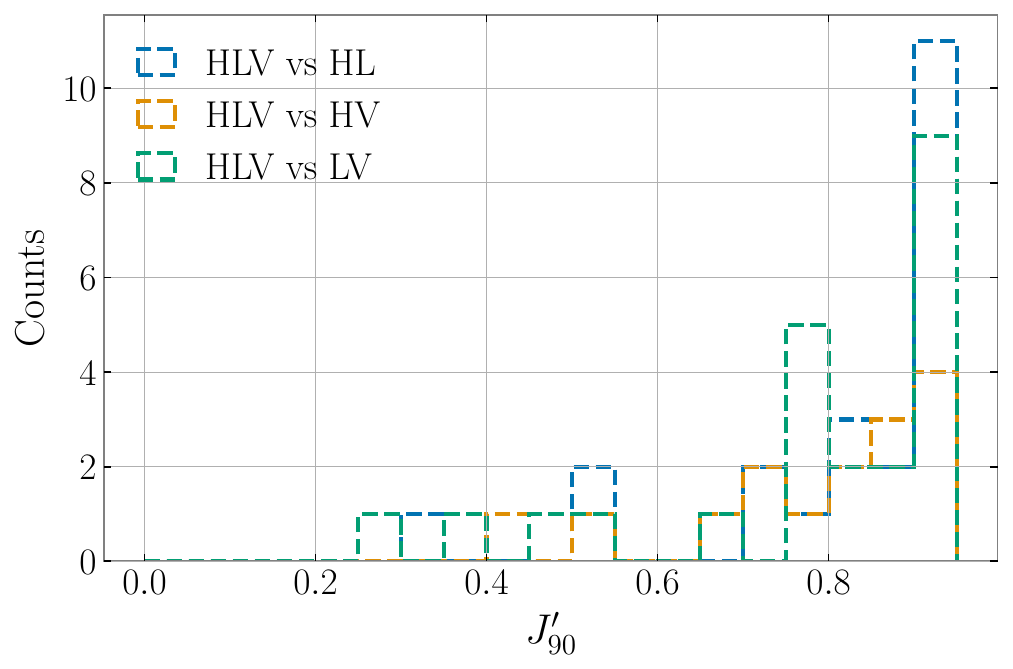}
    \caption{Distribution of renormalized Jaccard index \( J'_{90} \) comparing HLV with HL, HV, and LV skymaps for O3 alerts.}
    \label{fig:skyloc_O3_j90}
\end{figure}

The overlap analysis in Fig.~\ref{fig:skyloc_O3_j90} finds \(J'_{90}>0\) for all alerts. Median values are \(0.87\pm0.03\) (HLV vs.\ HL), \(0.90\pm0.02\) (HLV vs.\ HV), and \(0.84\pm0.03\) (HLV vs.\ LV), consistent with the injection benchmark (median \(\sim0.89\) for simulatons with real strain data in V1). KS tests comparing the O3 distributions with those from simulations (Table~\ref{tab:skyloc_O3_ks}) yield moderate \(D_{\mathrm{KS}}\) and \(p>0.05\) for all clean–strain comparisons, indicating no statistically significant differences. Results are similar when compared to injections using real strain data in Virgo, except for HLV vs.\ HL where \(p=0.041\) suggests a small mismatch, plausibly associated with low–significance transient artifacts in V1.

\begin{table}[htbp]
    \centering
    \caption{Kolmogorov--Smirnov test comparing O3 alerts and BBH injection distributions of \( J'_{90} \).}
    \label{tab:skyloc_O3_ks}
    \begin{tabular}{lcccc}
        \hline
        Comparison & \multicolumn{2}{c}{Gaussian-stationary noise} & \multicolumn{2}{c}{Online $h(t)$} \\
        & \( D_{\mathrm{KS}} \) & $p_{\mathrm{KS}}$ & \( D_{\mathrm{KS}} \) & $p_{\mathrm{KS}}$ \\
        \hline\hline
        HLV vs. HL & 0.225 & 0.055 & 0.234 & 0.041 \\
        HLV vs. HV & 0.153 & 0.372 & 0.153 & 0.368 \\
        HLV vs. LV & 0.155 & 0.356 & 0.164 & 0.290 \\
        \hline
    \end{tabular}
\end{table}

\subsection{Outlier Cases}
\label{sec:skyloc_special_cases_O3}

A small subset of O3 alerts exhibits distinct posterior behavior -- either pronounced KL–divergence asymmetries or reduced skymap overlap. These cases are identified in \S~\ref{sec:skyloc_consistency_O3} and summarized in Table~\ref{tab:skyloc_O3_j90_outliers}, which lists per–detector SNRs and \(J'_{90}\) for HLV relative to each two–detector baseline. Within this set, two alerts (\texttt{S190510g}, \texttt{S190718y}) have larger HLV 90\% credible areas than HL. Among the five, two alerts (\texttt{S200316bj}, \texttt{S200208q}) display Virgo–induced posterior shifts and sharpening consistent with the injection study, whereas three (\texttt{S200213t}, \texttt{S190510g}, \texttt{S190718y}) show patterns compatible with non–astrophysical origin or glitch contamination. 

\begin{table*}[htbp]
    \centering
    \caption{Selected O3 events with notable posterior shifts or low contour overlap. \(p_{\mathrm{astro}}\) is the astrophysical probability.}
    \label{tab:skyloc_O3_j90_outliers}
    \begin{tabular}{lccccccc}
        \hline
        GW Event & $\rho_H$ & $\rho_L$ & $\rho_V$ & \( J'_{90} \)(HLV vs. HL) & \( J'_{90} \)(HLV vs. HV) & \( J'_{90} \)(HLV vs. LV) & $p_{\mathrm{astro}}$ \\
        \hline\hline
        \texttt{S190510g}\tablefootmark{1}   & 2.84 & 8.30 & 4.66 & 0.33 & 0.41 & 0.79 & 42\% \\
        \texttt{S190718y}\tablefootmark{2}   & 7.95 & 8.49 & 2.97 & 0.51 & 0.98 & 0.26 & 2\% \\
        \texttt{S200208q}\tablefootmark{3}   & 6.38 & 7.33 & 4.23 & 0.84 & 0.98 & 0.92 & $>\!99\%$ \\
        \texttt{S200213t}\tablefootmark{4}   & 7.99 & 2.62 & 3.35 & 0.72 & 0.54 & 0.91 & 63\% \\
        \texttt{S200316bj}\tablefootmark{5}  & 5.39 & 8.09 & 3.12 & 0.94 & 1.00 & 0.96 & $>\!99\%$ \\
        \hline
    \end{tabular}
    \tablefoot{
      GraceDB links:
      \tablefoottext{1}{\href{https://gracedb.ligo.org/superevents/S190510g/}{S190510g}};
      \tablefoottext{2}{\href{https://gracedb.ligo.org/superevents/S190718y/}{S190718y}};
      \tablefoottext{3}{\href{https://gracedb.ligo.org/superevents/S200208q/}{S200208q}};
      \tablefoottext{4}{\href{https://gracedb.ligo.org/superevents/S200213t/}{S200213t}};
      \tablefoottext{5}{\href{https://gracedb.ligo.org/superevents/S200316bj/}{S200316bj}}.}
\end{table*}

For \texttt{S200316bj}, the LIGO SNRs are moderate (\(\rho_H=5.4\), \(\rho_L=8.1\)) and Virgo contributes \(\rho_V=3.1\). The HLV sky area is smaller than in all two–detector baselines, indicating improved localization. However, \(J'_{90}\) is lower for HLV vs.\ HL (0.94) than for HLV vs.\ HV (1.00) and HLV vs.\ LV (0.96), implying that Virgo provides additional directional information that shifts the HL posterior while reducing the HLV sky area. \texttt{S200208q} is analogous, with \(\rho_V=4.23\): the HLV posterior is sharper and displaced relative to HL, as indicated by \(J'_{90}(\mathrm{HLV}\text{ vs.\ HL})=0.84\). Both events have $p_{\rm astro}>\!99\%$ and are reported in GWTC--3.

By contrast, \texttt{S200213t} has low \(\rho_L=2.62\), higher \(\rho_H=7.99\), and \(\rho_V=3.35\). Overlap is poor for HLV vs.\ HL and HLV vs.\ HV (\(J'_{90}=0.72\) and \(0.54\), respectively) but good for HLV vs.\ LV (\(0.91\)). This indicates that new information from both L1 and V1 significantly shifts the corresponding two–detector baselines despite their low SNRs relative to H1. Such behavior -- when the added LIGO detector has very low SNR (here L1) -- is not expected from the injection study and points to inconsistencies between two–detector contours, suggesting a spurious noise contribution in L1. The candidate held a $p_{\rm astro}$ value of $63\%$ but was later excluded from GWTC–3 as likely non–astrophysical following offline analyses. Similarly, \texttt{S190510g} is dominated by L1 with very low H1 SNR (\(\rho_H=2.84\), \(\rho_L=8.30\), \(\rho_V=4.66\)), and overlaps between HLV and all two–detector baselines are weak; the event was detected only by \texttt{GstLAL} in low-latency CBC searches and deemed non–significant in offline analysis. \texttt{S190718y} has balanced LIGO SNRs (\(\rho_H\!\approx\!\rho_L\!\approx\!8\)) and moderate \(\rho_V=2.97\); both H1 and V1 provide low support for their respective two–detector baselines (LV and HL), as \(J'\) values are small. A glitch \(\sim40\) s before merger in L1 was also identified, and the event was later classified as non–astrophysical and not included in GWTC--3. Consistently, both \texttt{S190510g} and \texttt{S190718y} were reported in low latency with \(p_{\mathrm{astro}}<50\%\).

\section{Conclusions}

The first direct GW detection from a BBH merger \texttt{GW150914} by LIGO opened the field of GW astronomy. The first BNS detection, \texttt{GW170817}, was observed jointly by LIGO and Virgo; coordinated electromagnetic follow–up led to the discovery of electromagnetic counterparts that enabled new tests of fundamental physics. Rapid sky localization of GW sources is therefore essential to enable prompt searches for electromagnetic and neutrino counterparts: candidate CBC events are identified by matched–filter search algorithms \texttt{BAYESTAR} produces skymaps within minutes that guide follow–up observations.

Motivated by this operational role -- and by Virgo’s lower sensitivity relative to the LIGO observatories -- our aim was to quantify the extent to which a third detector improves low-latency sky localization. We generated simulated GW signals from three CBC populations -- BBH, NSBH, and BNS -- and produced rapid \texttt{BAYESTAR} skymaps using the \texttt{pycbc\_make\_skymap} package. Because real interferometer data contain noise artifacts (glitches and spectral lines) that can perturb signals and distort skymaps, we adopted a two-step design. First, we established an idealized baseline by injecting the same event set into stationary Gaussian noise for Hanford (H1), Livingston (L1), and Virgo (V1), and assessed localization across detector configurations. Second, holding the injection set fixed event-by-event, we repeated the analysis after replacing only Virgo’s simulated noise with real O3b Virgo strain (while keeping H1 and L1 Gaussian). This isolates Virgo’s effect in real data conditions and low-SNR -- where the signal can approach the noise floor.

To assess statistical calibration we used  PP plots. To compare three–detector (HLV) skymaps with their two–detector counterparts, we then employed three complementary metrics: (i) the \emph{90\% CRs sky–area ratio} $A^{90}_{\mathrm{HLV}}/A^{90}_{\mathrm{2\text{-}det}}$ to summarize sharpening; (ii) the \emph{Kullback–Leibler divergences} $D_{\mathrm{KL}}(\text{HLV}\,\|\,\text{2-det})$ and $D_{\mathrm{KL}}(\text{2-det}\,\|\,\text{HLV})$, whose directionality indicates whether HLV concentrates probability within the two–detector support or shifts it elsewhere; and (iii) the \emph{renormalized Jaccard index} $J'_{90}$, which measures the overlap of the $90\%$ CR between configurations.

PP–plots show systematic undercoverage (overconfident posteriors) across all configurations, with larger deviations at higher total mass and for the two–detector HL configuration. Calibration is best for BNS: for configurations that include Virgo (HV, LV, HLV), PP–plot deviations for BNS remain within $3\sigma$, indicating a minimal risk of counterpart loss due to mislocalization in BNS detections. The origin of skymap posterior overconfidence is left for further detailed investigation.

The diagnostics separate three behaviors. (i) \emph{Sharpening}: for HLV versus HV or LV with a moderately informative added detector, we find reduced $A^{90}$, high overlap ($J'_{90}\!\approx\!1$), and the characteristic KL asymmetry $D_{\mathrm{KL}}(\text{2-det}\,\|\,\text{HLV}) > D_{\mathrm{KL}}(\text{HLV}\,\|\,\text{2-det})$, indicating that HLV concentrates probability within the support of the two–detector map. (ii) \emph{Repositioning}: when Virgo’s response is strong, posterior support can shift; for HLV versus HL the KL asymmetry can reverse and $J'_{90}$ drops to zero more frequently, showing that HLV does not always reinforce HL but can correct it by moving probability to the region jointly supported by all three detectors. Additionally, in high–SNR near-symmetric two-LIGO configurations, the added LIGO detector reduces the $90\%$ sky area but the reduction ratio saturates. (iii) \emph{Degradation}: at low Virgo significance ($\rho_V<5$), the HLV searched area worsens compared to HL's; this occurs in $14.0\%$ of cases with simulated Gaussian noise and $19.8\%$ with real O3b Virgo data.

From these patterns we identify practical warning signs of skymap inconsistency: simultaneously low $J'_{90}$ in more than one HLV vs. 2-det comparison; a reversal of the expected KL asymmetry (opposite to sharpening) for HLV versus LV or HV; and cases in which $A^{90}$ increases after adding a detector. Applying this framework to O3 public alerts, we observe the same behaviors as in simulations: events with moderate Virgo SNR exhibit the expected sharpening and high overlap, whereas alerts with low overlap or unexpected asymmetric KL divergences correspond to known or suspected issues in data quality or marginal signal significance highlighting potential non-astrophysical origin of the event. 

These considerations extend to downstream offline analyses. \texttt{BAYESTAR} also returns a luminosity–distance estimate together with the 2D skymap; while the localization volume is expected to improve with the addition of a third detector \citep{DelPozzo:2018dpu}, we did not assess its accuracy here, which warrants a dedicated study. Sky localization further enters galaxy–catalog \textit{dark siren} cosmology, where GW sky–distance posteriors are combined with galaxy redshift catalogs (e.g., \texttt{GLADE+} \citep{Dalya:2021ewn}) to infer the Hubble constant $H_0$ and properties of compact object populations such as source–frame mass distributions and merger rates \citep{LIGOScientific:2021aug}. Current implementations in the \texttt{GWCosmo} \citep{Gray:2023wgj} and \texttt{ICAROGW} \citep{Mastrogiovanni:2023zbw} pipelines typically rely on full parameter–estimation (PE) posteriors (e.g., from \texttt{Bilby} \citep{Ashton2019_Bilby, Morisaki:2023kuq}) rather than \texttt{BAYESTAR}, which provide sky localization a few hours after the GW detection. Previous studies have evaluated sky localization performance using full PE \citep{Berry:2014jja, Emma:2024mjs}; in particular, \S4.2 of \citet{Berry:2014jja} reports good agreement between \texttt{BAYESTAR} and PE skymaps. In contrast, the O4 low-latency MDC \citep{Chaudhary:2023vec} shows a comparison with \texttt{Bilby} for BNS signals, where it demonstrates a better calibration ($< 3\sigma$) and \texttt{Bilby} skymaps yield searched areas smaller by approximately a factor of two relative to \texttt{BAYESTAR}. Extending our investigation to PE skymaps to quantify how localization overconfidence propagate into catalog–based cosmology is therefore an essential next step, since systematic mislocalization or distance misestimation can bias cosmological and population inferences. Finally, we did not investigate the impact of strain–calibration uncertainties on skymap systematics, nor potential biases that could be introduced by matched–filter settings in the search pipelines (e.g. different noise ASD estimation procedures or data sampling rates). Such choices can induce small changes in SNR and arrival time estimates for the same event. Calibration effects, in particular, may become significant for high–SNR events with tight sky localizations.

Looking ahead, the next era of GW astronomy with third-generation detectors -- Einstein Telescope \citep{Abac:2025saz} and Cosmic Explorer \citep{Hall:2022dik} -- is expected to raise the number of CBC detections to $\mathcal{O}(10^4)$ per observing run within the next decade \citep{Chen:2024gdn}. Potential biases that are negligible at today’s $\mathcal{O}(10^2)$ events scale will become statistically significant. Priorities therefore include improved uncertainty modeling in \texttt{BAYESTAR} and/or targeted, mass- and network-dependent calibration (e.g., introducing a likelihood scaling factor $\xi<1$), together with automated skymap–reliability indicators accompanying GW alerts. These measures will enhance the fidelity of low-latency skymaps and strengthen multi-messenger astronomy with GW observations.

\begin{acknowledgements}
We acknowledge the LIGO–Virgo–KAGRA Collaboration and thank the detector–characterization teams for their essential work. We are particularly grateful to Tito Dal Canton for his guidance and technical insights throughout this study. We also thank Florent Robinet, S\'ebastien Viret, Ines Bentara, Leo Singer and Pierre-Alexandre Duverne for their helpful input. This work used computational resources provided by the LIGO Laboratory and supported by the U.S.\ National Science Foundation (NSF) under Grants PHY–0757058 and PHY–0823459, and by the Computing Center of the \emph{Institut National de Physique Nucl\'eaire et de Physique des Particules} (CC–IN2P3/CNRS). This research has made use of data and/or software obtained from the Gravitational Wave Open Science Center (gwosc.org), a service of the LIGO Scientific Collaboration, the Virgo Collaboration, and KAGRA.

This material is based upon work supported by NSF’s LIGO Laboratory, a major facility fully funded by the NSF; by the Science and Technology Facilities Council (STFC) of the United Kingdom; by the Max Planck Society (MPS) and the State of Niedersachsen, Germany, for the construction of Advanced LIGO and the construction and operation of the GEO600 detector; and by the Australian Research Council for additional support to Advanced LIGO. Virgo is funded, through the European Gravitational Observatory (EGO), by the French \emph{Centre National de la Recherche Scientifique} (CNRS), the Italian \emph{Istituto Nazionale di Fisica Nucleare} (INFN), and the Dutch Nikhef, with contributions from institutions in Belgium, Germany, Greece, Hungary, Ireland, Japan, Monaco, Poland, Portugal, and Spain. KAGRA is supported by the Ministry of Education, Culture, Sports, Science and Technology (MEXT) and the Japan Society for the Promotion of Science (JSPS) in Japan; by the National Research Foundation (NRF) and the Ministry of Science and ICT (MSIT) in Korea; and by Academia Sinica (AS) and the National Science and Technology Council (NSTC) in Taiwan.

\end{acknowledgements}

\bibliographystyle{aa}
\bibliography{main}

\end{document}